\documentclass[10pt,twocolumn,twoside]{IEEEtran}
\usepackage{amsmath,amssymb,bm,cite,algorithm,algpseudocode,amsthm}
\usepackage{graphicx, xcolor}
\usepackage{bm}
\usepackage{acronym}
\usepackage{booktabs}
\usepackage{tabularx}

\newacro{ML}{Maximum Likelihood}
\newacro{LLF}{Log-Likelihood Function}
\newacro{SS-SW-OMP+Th}{Subcarrier Selection - Simultaneous Weighted - Orthogonal Matching Pursuit + Thresholding}
\newacro{SC-SS-SIMGW-IPM}{Spatially Consistent - Subcarrier Selection - Simultaneous Iterative Multi Gradient Weighted - Iterative Projection Maximization}
\newacro{SC-SS-SW-IPM}{Spatially Consistent - Subcarrier Selection - Simultaneous Weighted - Iterative Projection Maximization}
\newacro{SS-SIGW-OLS}{Subcarrier Selection - Simultaneous Iterative Gridless Weighted - Orthogonal Least Squares}
\newacro{SS-SW-ROLS}{Subcarrier Selection - Simultaneous Weighted - Reduced Orthogonal Least Squares}
\newacro{AM}{Alternating Maximization}
\newacro{CSI}{Channel State Information}
\newacro{TTI}{Transmission Time Interval}
\newacro{AoA}{Angle-of-Arrival}
\newacro{AoD}{Angle-of-Departure}
\newacro{LSP}{Large-Scale Parameters}
\newacro{SSP}{Small-Scale Parameters}
\newacro{DS}{Delay Spread}
\newacro{AS}{Angular Spread}
\newacro{mmWave}{Millimeter Wave}
\newacro{LOS}{Line-of-Sight}
\newacro{NLOS}{Non Line-of-Sight}
\newacro{CRLB}{Cram\'{e}r-Rao Lower Bound}
\newacro{NR}{New Radio}
\newacro{MU}{Multi-User}
\newacro{UMi}{Urban Micro Cell}
\newacro{NMSE}{Normalized Mean Squared Error}
\newacro{MIMO}{Multiple-Input Multiple-Output}
\newacro{MMSE}{Minimum Mean Square Error}
\newacro{MVUE}{Minimum Variance Unbiased Estimator}
\newacro{KF}{Kalman Filter}
\newacro{PF}{Particle Filter}
\newacro{MPF}{Marginalized Particle Filter}
\newacro{GMPF}{Generalized Marginalized Particle Filter}
\newacro{MT}{Mobile Terminal}

\usepackage{pifont}

\newcommand{\SNR}{\mathrm{SNR}}
\newcommand{\vect}{\mathrm{vec}}
\newcommand{\unvect}{\mathrm{unvec}}
\newcommand{\bsfCw}{\bsfC_\mathrm{w}}

\newcommand{\blkdiag}{\mathrm{blkdiag}}
\newcommand{\blkrow}{\mathrm{blkrow}}
\newcommand{\diag}{\mathrm{diag}}
\newcommand{\trace}{\mathrm{trace}}
\newcommand{\rank}{\mathrm{rank}}
\newcommand{\Nr}{N_{\mathrm{r}}}
\newcommand{\Nt}{N_{\mathrm{t}}}
\newcommand{\Lt}{L_{\mathrm{t}}}
\newcommand{\Lr}{L_{\mathrm{r}}}
\newcommand{\Ns}{N_{\mathrm{s}}}
\newcommand{\Mtck}{M_{\mathrm{tck}}}

\newcommand{\Ts}{T_{\mathrm{s}}}

\allowdisplaybreaks

\def\ba{{\mathbf{a}}}

\def\bx{{\mathbf{x}}}

\def\b0{{\mathbf{0}}}

\def\bA{{\mathbf{A}}}

\def\bC{{\mathbf{C}}}
\def\bD{{\mathbf{D}}}

\def\bF{{\mathbf{F}}}

\def\bH{{\mathbf{H}}}
\def\bI{{\mathbf{I}}}

\def\bM{{\mathbf{M}}}

\def\bP{{\mathbf{P}}}

\def\bW{{\mathbf{W}}}
\def\bX{{\mathbf{X}}}

\def\bsfA{\bm{\mathsf{A}}}
\def\bsfB{\bm{\mathsf{B}}}
\def\bsfC{\bm{\mathsf{C}}}
\def\bsfD{\bm{\mathsf{D}}}

\def\bsfF{\bm{\mathsf{F}}}
\def\bsfG{\bm{\mathsf{G}}}
\def\bsfH{\bm{\mathsf{H}}}

\def\bsfK{\bm{\mathsf{K}}}

\def\bsfP{\bm{\mathsf{P}}}
\def\bsfQ{\bm{\mathsf{Q}}}
\def\bsfR{\bm{\mathsf{R}}}

\def\bsfT{\bm{\mathsf{T}}}
\def\bsfU{\bm{\mathsf{U}}}
\def\bsfV{\bm{\mathsf{V}}}
\def\bsfW{\bm{\mathsf{W}}}
\def\bsfX{\bm{\mathsf{X}}}
\def\bsfY{\bm{\mathsf{Y}}}

\def\sfj{{\mathsf{j}}}

\def\sfs{{\mathsf{s}}}

\def\sf0{{\mathsf{0}}}

\def\bsfa{{\bm{\mathsf{a}}}}

\def\bsfee{{\bm{\mathsf{e}}}}

\def\bsfg{{\bm{\mathsf{g}}}}
\def\bsfh{{\bm{\mathsf{h}}}}

\def\bsfn{{\bm{\mathsf{n}}}}

\def\bsfq{{\bm{\mathsf{q}}}}
\def\bsfr{{\bm{\mathsf{r}}}}
\def\bsfs{{\bm{\mathsf{s}}}}

\def\bsfx{{\bm{\mathsf{x}}}}
\def\bsfy{{\bm{\mathsf{y}}}}
\def\bsfz{{\bm{\mathsf{z}}}}
\def\bsf0{{\bm{\mathsf{0}}}}

\begin{document}
\title{Channel Tracking and Hybrid Precoding for Wideband Hybrid Millimeter Wave MIMO Systems}
\author{Javier Rodr\'{i}guez-Fern\'{a}ndez$^{\dag}$, Nuria Gonz\'{a}lez-Prelcic$^{\dag}$ and Takayuki Shimizu$^{\ddag}$ \\ \thanks{This work was supported in part by the National Science Foundation under Grant No. CNS-1702800 and by a gift from Toyota InfoTechnology Center.}
$^\dag$ The University of Texas at Austin, Email: $\{$javi.rf,ngprelcic$\}$@utexas.edu\\
$^\ddag$ Toyota InfoTechnology Center, USA, Inc. Email: tshimizu@us.toyota-itc.com
}

\maketitle

\begin{abstract}
 A major source of difficulty when operating with large arrays at \ac{mmWave} frequencies is to estimate the wideband channel, since the use of hybrid architectures acts as a compression stage for the received signal. Moreover, the channel has to be tracked and the antenna arrays regularly reconfigured to obtain appropriate beamforming gains when a mobile setting is considered. In this paper, we focus on the problem of channel tracking for frequency-selective \ac{mmWave} channels, and propose two novel channel tracking algorithms that leverage prior statistical information on the angles-of-arrival and angles-of-departure. Exploiting this prior information, we also propose a precoding and combining design method to increase the received SNR during channel tracking, such that near-optimum data rates can be obtained with low-overhead. In our numerical results, we analyze the performance of our proposed algorithms for different system parameters. Simulation results show that, using channel realizations extracted from the 5G New Radio channel model, our proposed channel tracking framework is able to achieve near-optimum data rates.
\end{abstract}


\section{Introduction}\label{sec:Intro}  

\begin{figure*}[t!]
\centering
\includegraphics[width=\textwidth]{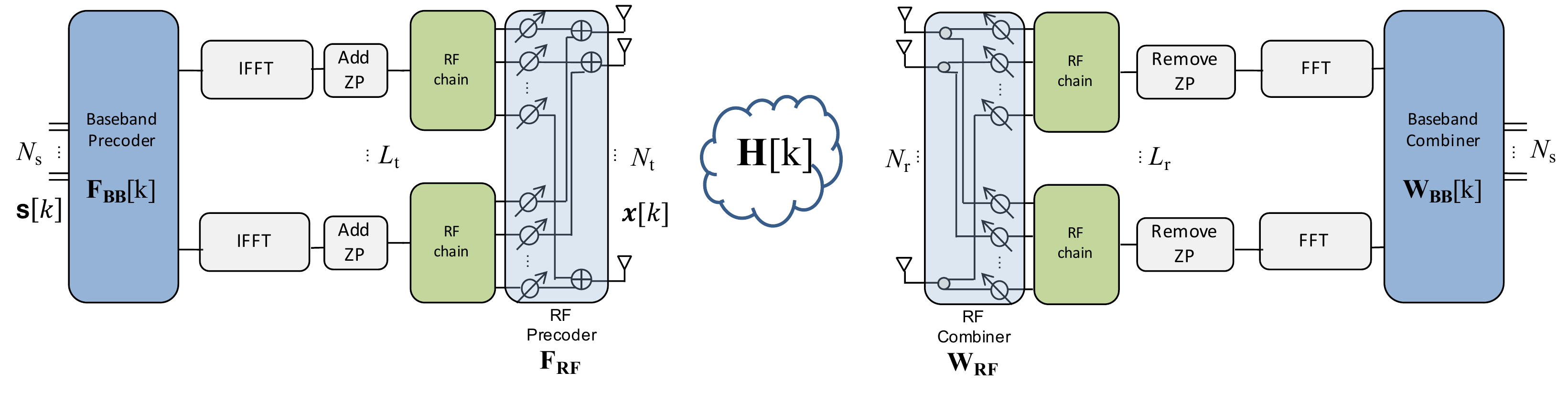} 
\caption{Illustration of the structure of a hybrid MIMO architecture, which include analog and digital precoders and combiners.}     
\label{fig:hybrid_architecture}        
\end{figure*}

MmWave \ac{MIMO} is a key ingredient for the fifth generation of wireless communications, which has been standardized as 5G \ac{NR} \cite{5G_standard}. 5G \ac{NR} considers an unified framework allowing an efficient use of sub-6 GHz and millimeter wave bands. The current version of the standard lacks, however, of an efficient procedure for configuring the antenna arrays. The proposed beam training and beam tracking protocol \cite{Modular-High-Resolution-5G} incurs in a high overhead, and it is not appropriate for high mobility scenarios \cite{mmWaveV2X}. 

Channel estimation for initial access is an effective alternative to beam training to configure mmWave antenna arrays. It has been studied in the literature for narrowband \cite{Malloy2012,Malloy2012a,Iwen2012, AlkAyaLeuHea:Channel-Estimation-and-Hybrid:14,Ramasamy2012a,Ramasamy2012b,Berraki2014,Lee2014,RiaRusPreAlkHea:Hybrid-MIMO-Architectures:16,Marzi2016:ChanEstTracking,ImprovedChEst,ChEstHybridPrecmmWave}, and frequency-selective channels \cite{JSAC_channel_est,RodGonVen:A-Frequency-Domain-Approach-to-Wideband:17,RodGonVenHea:TWC_18,LiuYangDingWangSong:2DStructuredCS,GaoDaiWan:Channel-estimation-mmWave-massiveMIMO:16,CAMSAP_2017_Offgrid,ComaRodGonCasHea:JSTSP:18}, yet little attention has been drawn towards the problem of channel tracking. An interesting feature of the new 5G \ac{NR} channel model \cite{5G_channel_model} is that it incorporates spatial consistency. This way, a model for the temporal evolution of different channel parameters could be leveraged, and effective strategies for channel tracking at mmWave frequencies could be devised.

The problem of channel tracking under a narrowband communication model has been considered in recent work \cite{Mobilize_mmWave_joint_beam_channel_tracking,Tracking-mmWave-channel-dynamics,George_tracking,mmWave-MIMO-channel-tracking-systems,Mobile-mmwave-acquisition-tracking-abrupt-change-detection,Priori-aided-channel-tracking,Sampling-based-tracking-mmwave,Tracking-abruptly-changing-channels-mmwave,Tracking-angles-departure-arrival}, although the mmWave channel is frequency-selective. Besides this unrealistic assumption, prior work on channel tracking considers different ad-hoc mathematical approaches to model the channel dynamics, that is, how the \ac{AoA}, \ac{AoD}, and channel gains evolve with time. For instance, prior work considering time correlation for the channel gains assumes a first-order Gauss-Markov process with very large correlation factor (i.e. $0.995$), and makes no distinction between the correlation of \ac{LOS} and \ac{NLOS} multipath components \cite{George_tracking}, \cite{mmWave-MIMO-channel-tracking-systems}, \cite{Beam-tracking-Vutha}, \cite{CAMSAP_Tracking}, while the spatial correlation of channel gains has been shown to be different for \ac{LOS} and \ac{NLOS} components \cite{Rap:17:Spatial_correlation}, the latter exhibiting lower correlation values. As to the \ac{AoA} and \ac{AoD}, prior work considers that the angles evolve according to either a Gaussian distribution (which is a mathematical artifact to enable the use of a linearized Kalman filter) \cite{Beam-tracking-Vutha}, \cite{Tracking-angles-departure-arrival}, \cite{Mobile-mmwave-acquisition-tracking-abrupt-change-detection}, a uniform distribution \cite{George_tracking}, \cite{mmWave-MIMO-channel-tracking-systems}, or deterministically \cite{CAMSAP_Tracking}, while these parameters have been shown to follow a Laplacian distribution \cite{Molisch:Statistical_mmWave_channel:17}. Finally, the  spatial consistency model proposed in 5G \ac{NR} differs in one or another sense from all these other models. 

To the best of our knowledge, there are only two papers dealing with the problem of \ac{mmWave} channel tracking for frequency-selective single-user scenarios \cite{CAMSAP_Tracking}, \cite{Asilomar17_Tracking}. The limitations of \cite{CAMSAP_Tracking} are:
\begin{itemize}
	\item The considered channel model is not band-limited, that is, the effect of the equivalent transmit-receive pulse-shaping and analog filtering is neglected, thus leading to both an unrealistic and sparser channel model. This makes it easier to both estimate and keep track of the \ac{mmWave} channel variations, as already discussed in \cite{RodGonVenHea:TWC_18}.
	\item A very high spatial correlation for the channel gains is assumed ($0.995$), which is very unlikely to hold under a tracking periodicity of $20$ ms, as considered in the simulation results of \cite{CAMSAP_Tracking}. Further, such high correlation value would be more realistic for a \ac{LOS} multipath component, but it is assumed to hold for the \ac{NLOS} components, contradicting the empirical results in \cite{Rap:17:Spatial_correlation}.
	\item No noise statistics in the received signal are employed, despite the use of more than a single RF chain.  As analyzed in \cite{RodGonVenHea:TWC_18}, both estimation performance and the final spectral efficiency can be improved when considering the statistical features of the noise.
	\item The baseline estimation method used for initial channel acquisition is assumed to exhibit estimation error equal to the \ac{CRLB}, which only could hold when the \ac{AoA} and \ac{AoD} fall within a quantized spatial grid, and both the number of subcarriers and number of training frames are large enough \cite{RodGonVenHea:TWC_18}. 
	\end{itemize}

In our prior work in \cite{Asilomar17_Tracking}, we developed a gradient algorithm for channel tracking that approximates the \ac{ML} estimator conditioned on a certain sparsity level, which was estimated during initial channel acquisition. Since the angular resolution of an $N$-element ideal antenna array is limited by $1/N$, two multipath components arriving with an angular separation smaller than this limit cannot be distinguished. Therefore, the \ac{ML} estimator cannot be found in general owing to clustering and the use of finite-resolution antenna arrays. The main limitations of \cite{Asilomar17_Tracking} are the high complexity and the still high tracking overhead, which is caused by the method used to design the hybrid precoders and combiners during tracking, and the introduction of hardware impairments in the simulations (cross polarization effects, miscalibration errors, beam-squint, and non-omnidirectional antenna response taken from the 3GPP 3D antenna array model \cite{5G_channel_model}).
\subsection{Contributions}

In this paper, we focus on the problem of channel tracking for frequency-selective mmWave \ac{MIMO} systems under the channel model used for 5G \ac{NR}. We focus on the single-user scenario with transceivers being equipped with hybrid \ac{MIMO} architectures \cite{Rial_switchOrShifter:Access2016}, although the proposed channel tracking framework can be extended to the \ac{MU} scenario similarly to \cite{ComaRodGonCasHea:JSTSP:18}. Further, we also focus on the problem of precoder and combiner design for channel tracking, whereby our interest is in achieving near-optimum data rates while keeping low overhead and low computational complexity as well. The detailed contributions of our paper are listed hereafter:

\begin{itemize}
	\item We formulate the channel tracking problem under a single-user frequency-selective \ac{mmWave} \ac{MIMO} channel scenario. We present a \textit{general} formulation of the channel tracking problem, in which prior statistical information on the different \ac{SSP} may or may not be available.
		
	\item We propose two novel algorithms to perform channel tracking at \ac{mmWave}. The first algorithm relies on an off-grid representation of the \ac{MIMO} channel, and it attempts to find the \textit{sparsity-constrained} \ac{ML} estimator of the \ac{mmWave} \ac{MIMO} channel. The second algorithm, however, aims at finding the \textit{sparsity-constrained} \ac{MMSE} estimator of the \ac{mmWave} \ac{MIMO} channel when prior information on the statistics of the \ac{SSP} is available. Thus, we present both \textit{classical} and \textit{Bayesian} solutions to the problem of wideband channel tracking, and compare their performance as a function of different system parameters to show which is the gain coming from considering prior statistical information. The proposed algorithms allow not only finding accurate estimates of the \ac{mmWave} \ac{MIMO} channel, but they also result in very low training overhead. Consequently, they enable us to obtain high spectral efficiency values, even when mobility is high and the distance between transmitter and receiver is small, thereby making the \ac{AoA} and \ac{AoD} harder to track.

	\item We propose a method to design hybrid precoders and combiners for channel tracking exploiting prior angular information from the channel estimates at the immediately preceding channel slot. This technique aims at minimizing the chordal distance between the hybrid precoders (combiners) and the right (left) channel singular subspaces, exploiting information from the estimated channel's left and right singular subspaces. Further, we prove that this method also maximizes the received $\SNR$ and, consequently, the Fisher Information on the \ac{AoA} and \ac{AoD}, thereby being asymptotically optimal for channel tracking.

	\item In our numerical results, unlike prior work, we do not assume that the \ac{MIMO} channel remains invariant during either initial channel estimation, channel tracking, or data transmission, so that we evaluate the robustness of the proposed channel tracking and hybrid precoding and combining algorithms under a realistic framework in which each transmitted OFDM symbol experiments a different \ac{mmWave} \ac{MIMO} channel. This is, to the best of our knowledge, the first paper that deals with this problem and does not assume a static channel during transmission of training and data symbols within a channel tracking slot. Therefore, \textit{the transmitter and receiver are not assumed to be synchronized with the channel dynamics}, what leads to highly robust channel tracking strategies. 
\end{itemize}

In the numerical results, we evaluate our proposed channel tracking and hybrid precoding algorithms, which achieve near-optimum data rates even when evaluated using realistic channel series obtained with the QuaDRiGa channel simulator \cite{QuaDRiGa_IEEE}, \cite{QuaDRiGa_Tech_Rep}, that implements the 5G NR channel model.
We also show a comparison between our classical and Bayesian strategies, to show the benefit of using prior statistical information to further reduce training overhead, and thereby increase the resulting spectral efficiency.

\textit{Notation:} Vectors and matrices are denoted by boldface lowercase letters and boldface capital letters, respectively. The conjugate, transpose, Hermitian transpose, and Moore-Penrose pseudoinverse of a matrix $\mathbf{A}$ are denoted by $\mathbf{A}^\text{C}$, $\mathbf{A}^{T}$, $\mathbf{A}^*$, and $\mathbf{A}^\dag$, respectively. The $n \times n$ identity matrix is denoted as $\mathbf{I}_{n}$ ($n\geq2$). $\|\mathbf{a}\|_p$ stands for the $p$-norm of $\mathbf{a}$, and ${\rm diag}\{\mathbf{a}\}$ denotes a square diagonal matrix with $\mathbf{a}$'s elements in its main diagonal. $[\mathbf{A}]_{i,j}$ represents $\mathbf{A}$'s $(i,j)$-th element and $i$-th column, respectively. $\mathbf{A}\circ\mathbf{B}$ represents the Khatri-Rao product of $\mathbf{A}$ and $\mathbf{B}$, while $\mathbf{A}\otimes\mathbf{B}$ denotes their Kronecker product. $\left|\mathcal{F}\right|$ is the cardinality of set $\mathcal{F}$ and $|\cdot|$ denotes the amplitude of a complex number. Notation $\bx\sim\mathcal{CN}\left(\bm 0,\bC\right)$ indicates that $x$ is a circularly-symmetric complex Gaussian random variable with zero mean and covariance matrix $\bC$. Time-domain vectors (matrices) are represented using $\bx[n]$ ($\bX[n]$), whilst frequency-domain vectors (matrices) are represented using $\bsfx[k]$ ($\bsfX[k]$). We use $\bm \mu_{\bsfx}$ to denote the mean of a vector $\bsfx$, and $\bsfC_{\bsfx\bsfy}$ to denote the covariance matrix between two random vectors $\bsfx$,$\bsfy$.

\section{System and Channel Models}\label{sec:System_Model}

In this section, we introduce the models and assumptions for the different blocks of the communication system considered in
this paper.
\subsection{System Model}\label{Sec:System_Model}

We consider a single-user MIMO-OFDM communication link between a transmitter and a receiver equipped with $\Nt$ and $\Nr$ antennas, and $\Lt$ and $\Lr$ RF chains. Both transceivers are assumed to use a fully-connected hybrid architecture, as described in \cite{Rial_switchOrShifter:Access2016} and shown in Fig. \ref{fig:hybrid_architecture}. We will use super-index $n$ to denote the $n$-th channel slot, also known as \ac{TTI} in cellular standards, which is defined as the time window during which the channel may be considered to be constant.

\begin{figure}
\centering
\includegraphics[width=\columnwidth]{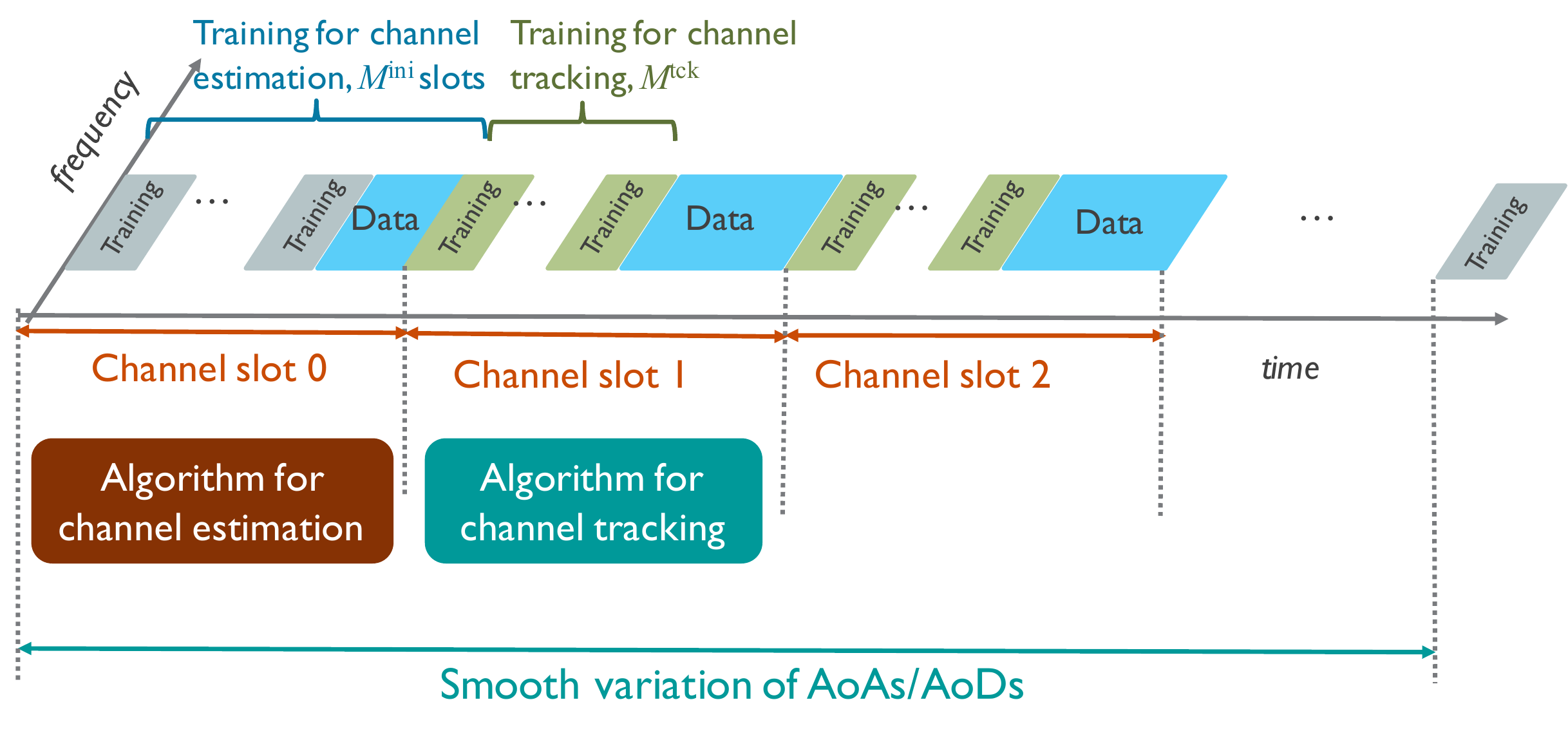}
\caption{Illustration of how \ac{mmWave} channel acquisition is obtained on a frame-by-frame basis, encompassing both a training phase and a data transmission phase.}
\label{fig:training_phases}
\end{figure}

Within a channel slot, there are two different transmission stages: i) a training phase to acquire \ac{CSI} in order to configure the hybrid antenna arrays, and ii) a data phase to communicate data vectors from the transmitter to the receiver and viceversa, as shown in Fig. \ref{fig:training_phases}. The first phase consists of the transmission of $M^{(n)}$ training symbols, during which control information is exchanged between transmitter and receiver in order to obtain \ac{CSI}. During initial acquisition ($n = 0$), the training phase lasts longer than during tracking ($n \geq 1$), owing to lack of prior information about the channel.
 In practical mmWave \ac{MIMO} links, variations are expected between two consecutive slots, and each slot should be long enough to accommodate both the training and data phases.  
 
During the data phase of the $n$-th channel slot, the transmitter uses a hybrid precoder $\bsfF^{(n)}[k] \in \mathbb{C}^{\Nt \times \Ns}$ for the $k$-th subcarrier, $k = 0,\ldots,K-1$, to transmit a $\Ns \times 1$ vector of data streams $\bsfs^{(n)}[k]$, in which $\Ns$ denotes the number of data streams to be transmitted. $\bsfF^{(n)}[k] = \bF_\text{RF}^{(n)}\bsfF_\text{BB}^{(n)}[k]$, with $\bF_\text{RF}^{(n)} \in \mathbb{C}^{\Nt \times \Lt}$ the analog precoder, common for all subcarriers, and $\bsfF_\text{BB}^{(n)}[k] \in \mathbb{C}^{\Lt \times \Ns}$ the baseband precoder for the $k$-th subcarrier. We denote the transmit power as $P_\text{tx}$, and we assume that $\sum_{k=0}^{K-1}\|\bsfF_\text{tr}^{(m,n)}[k]\|_F^2 = P_\text{tx}$. Likewise, the receiver uses a hybrid combiner $\bsfW^{(n)}[k] \in \mathbb{C}^{\Nr \times \Ns}$, $\bsfW^{(n)}[k] = \bW_\text{RF}^{(n)} \bsfW_\text{BB}^{(n)}[k]$, with $\bW_\text{RF}^{(n)} \in \mathbb{C}^{\Nr \times \Lr}$ the analog combiner, and $\bsfW_\text{BB}^{(n)}[k] \in \mathbb{C}^{\Lr \times \Ns}$ the baseband combiner at the $k$-th subcarrier. If the mmWave \ac{MIMO} channel for a given subcarrier $k$ and slot $n$ is denoted as $\bsfH^{(n)}[k]$, the received signal during transmission of a data stream $\bsfs^{(n)}[k] \in \mathbb{C}^{\Ns \times 1}$ is given by
\begin{equation}
\bsfy^{(n)}[k] = \bsfW_\text{BB}^{(n)*}[k]\bsfW_\text{RF}^{(n)*} \bsfH^{(n)}[k] \bsfF_\text{RF}^{(n)} \bsfF_\text{BB}^{(n)}[k] \bsfs^{(n)}[k] + \bsfn^{(n)}[k],
\end{equation}
where the $\Ns \times 1$ vector $\bsfn^{(n)}[k]$ is the complex-valued additive Gaussian-distributed received noise $\bsfn^{(n)}[k] \sim {\cal N}(\b0,\sigma^2 \bsfCw^{(n)}[k])$, with $\bsfCw^{(n)}[k]$ the noise covariance matrix given by $\bsfCw^{(n)}[k] = \bsfW_\text{BB}^{(n)*}[k]\bsfW_\text{RF}^{(n)*}\bsfW_\text{RF}^{(n)}\bsfW_\text{BB}^{(n)}[k]$. 

For the training phase, we need to sound the channel with a different combination of training precoder and combiner for every OFDM pilot.
This way, the transmitter uses a hybrid precoder $\bsfF_\text{tr}^{(m,n)}[k] \in \mathbb{C}^{\Nt \times \Lt}$ to transmit the $m$-th OFDM pilot. Such a precoder can be expressed as $\bsfF_\text{tr}^{(m,n)}[k] = \bF_\text{tr,RF}^{(m,n)}\bsfF_\text{tr,BB}^{(m,n)}[k]$, with $\bF_\text{tr,RF}^{(m,n)} \in \mathbb{C}^{\Nt \times \Lt}$ the training analog precoder, common for all subcarriers, and $\bsfF_\text{tr,BB}^{(m,n)}[k] \in \mathbb{C}^{\Lt \times \Lt}$ the training baseband precoder, which can be different for each subcarrier. We assume again that $\sum_{k=0}^{K-1}\|\bsfF_\text{tr}^{(m,n)}[k]\|_F^2 = P_\text{tx}$. Similarly to the transmitter, the training combiner can be expressed as $\bsfW_\text{tr}^{(m,n)}[k] = \bW_\text{tr,RF}^{(m,n)} \bsfW_\text{tr,BB}^{(m,n)}[k]$, with $\bW_\text{tr,RF}^{(m,n)} \in \mathbb{C}^{\Nr \times \Lr}$ the training analog combiner, and $\bsfW_\text{tr,BB}^{(m,n)}[k] \in \mathbb{C}^{\Lr \times \Lr}$ the training baseband combiner at the $k$-th subcarrier.  The received signal during training is given by
\begin{equation}
\begin{split}
\bsfy^{(m,n)}_\text{tr}[k] &= \bsfW_\text{tr}^{(m,n)*}[k] \bsfH^n[k] \bsfF_\text{tr}^{(m,n)}[k] \bsfs_\text{tr}^{(m,n)}[k] \\ &+ \bsfn^{(m,n)}[k],
\end{split}
\label{eq:rxtraining}
\end{equation}
where the $\Lr \times 1$ noise vector $\bsfn^{(m,n)}[k]$ follows the same distribution as in the data phase. 

\subsection{Frequency-Selective Channel Model}\label{Sec:Channel_Model}
The $d$-th delay tap of the mmWave \ac{MIMO} channel matrix corresponding to the $n$-th channel slot is defined using a clustered channel model \cite{schniter_sparseway:2014} as follows. Let $D \in \mathbb{N}$ be the channel delay tap length, $\alpha_{c,r}^{(n)} \in \mathbb{C}$ be the complex gain of the $r$-th ray within the $c$-th cluster, $\phi_{c,r}^n$, $\theta_{c,r}^n\in \mathbb{R}$ be the AoA and AoD, $\tau_{c,r}^n \in \mathbb{R}$ be the time-delay, and $p(\tau)$ be the equivalent response of transmit and receive pulse-shapes and other analog filtering evaluated at $\tau$. The total number of clusters is denoted by $C \in \mathbb{N}$, and the $c$-th cluster consists of $R_c \in \mathbb{N}$ rays. Last, let $\ba_\text{T}\left(\theta_{c,r}^n\right) \in \mathbb{C}^{\Nt \times 1}$ and $\ba_\text{R}\left(\phi_{c,r}^n\right) \in \mathbb{C}^{\Nr \times 1}$ be the transmit and receive array steering vectors evaluated on the \ac{AoD} and \ac{AoA} of each corresponding path. Then, the $\Nr \times \Nt$ channel matrix at delay tap $d = 0,\ldots,D-1$ is given by \cite{schniter_sparseway:2014}
\begin{equation}
\begin{split}
	\bH^n[d] &= \sum_{c=1}^{C}\sum_{r=1}^{R_c}\alpha_{c,r}^n p\left(d\Ts - \tau_{c,r}^n\right) \times \\ &\ba_\text{R}\left(\phi_{c,r}^n\right) \ba_\text{T}^*\left(\theta_{c,r}^n\right).
\end{split}
	\label{equation:channel_delay_tap}
\end{equation}
From \eqref{equation:channel_delay_tap}, we may express the \ac{MIMO} channel in the frequency domain, whereby taking the $K$-point DFT of \eqref{equation:channel_delay_tap} yields
\begin{equation}
\begin{split}
	\bsfH^n[k] &= \sum_{d=0}^{D-1}{\bH^n[d] e^{-j\frac{2\pi k d}{K}}} \\
	&= \bsfA_\text{R}\left(\bm \phi^n\right) \bsfG^n[k] \bsfA_\text{T}^*\left(\bm \theta^n\right),
\end{split}
	\label{equation:compact_freq_channel}
\end{equation}
where $\bA_\text{T}\left(\bm \theta^n\right) \in \mathbb{C}^{\Nt \times \sum_{c=1}^{C}{R_c}}$, $\bA_\text{R}\left(\bm \phi^n\right) \in \mathbb{C}^{\Nr \times \sum_{c=1}^{C}{R_c}}$ denote the transmit and receive antenna array responses, and $\bsfG^n[k] \in \mathbb{C}^{\sum_{c=1}^{C}{R_c} \times \sum_{c=1}^{C}{R_c}}$ is the diagonal matrix containing the complex channel gains.

Although the equations above will be used to derive the channel tracking algorithms proposed in this paper, we will use channel realizations generated with QuaDRiGa to test our tracking strategies, as described in Section~\ref{sec:Results}. For continuous-time evolution of the channel matrix, the feature of spatial consistency will be used, which is currently implemented in QuaDRiGa v2.0. This method updates \ac{SSP}s (delays, angles, power and phase of each channel path) as described in Section 7.6.3.2 in \cite{5G_channel_model} (Option B). This procedure makes generated paths not to change abruptly when the MT moves. To obtain this effect and generate a continuous time-series, the complete channel is splitted into several segments, for which \ac{MT}'s positions are updated, as well as both the \ac{SSP}s and \ac{LSP}s. Finally, the individual channel responses for each segment are combined such that a smooth channel response is obtained.

\section{Channel Tracking Algorithms}

In this section, we propose  novel algorithms for tracking the variations of the mmWave \ac{MIMO} channel.
The main assumptions we make regarding channel variations and tracking are summarized as follows:
\begin{itemize}	
	\item We assume that the number of paths is invariant along successive slots, i.e., there are no abrupt changes and only tracking is performed. If cluster blockage is present, the system needs to fully re-estimate the channel again. Future work will also consider the blockage dynamics. In this manuscript, we only focus on solving the problem of configuring the transmit and receive antenna arrays during channel slots that meet the spatial consistency model defined in \cite{5G_channel_model}.
	
	\item We assume that the \ac{AoA}s and \ac{AoD}s vary relatively smoothly between any two consecutively slots, such that estimates of these parameters obtained during the $(n-1)$-th slot can be used as prior information to refine those at the $n$-th slot. This assumption is reasonable if the \ac{MT} is not moving extremely fast (i.e. less than $400$ km/h).
	\end{itemize}

\subsection{Problem formulation}
During the training phase of the $n$-th channel slot, $M^{(n)}$ training frames are forwarded to the receiver. For simplicity, we assume $M^{(n)} = M_\text{tck}$, $\forall n \geq 1$. During the training phase, transmitter and receiver are assumed to use frequency-flat hybrid precoders and combiners. The reasons for this choice are two-fold: i) to compute the channel gains, matrix inverses (pseudoinverses) need to be computed at the receiver, so that the use of frequency-flat precoders and combiners allow computing these gains by using a single matrix inverse, thereby reducing computational complexity during channel tracking, and ii) it reduces memory storage needs at both transmitter and receiver.

The received signal during training in \eqref{eq:rxtraining} can be rewritten as
\begin{equation}
\begin{split}
	\bsfy^{(m,n)}_\text{tr}[k] &= \underbrace{\left(\bsfs^{(m,n)T}_\text{tr}[k]\bF_\text{tr}^{(m,n)T} \otimes \bW_\text{tr}^{(m,n)*}\right)}_{\bm \Phi^{(m,n)}[k]} \vect\{\bsfH^{(m,n)}[k]\} \\ &+ \bsfn^{(m,n)}[k].
\end{split}
	\label{equation:vect_signal_model_single_tck}
\end{equation}
Now, owing to dimensionality reduction at the RF stage of both transmitter and receiver, it is necessary to acquire several measurements to track the variations of the mmWave frequency-selective channel. Hence, defining $\bsfy^{(n)}[k] \triangleq [\bsfy^{(1,n)T}_\text{tr}[k],\ldots,\bsfy^{(M_\text{tck},n)T}_\text{tr}[k]]^T$, $\bm \Phi^{(n)}[k] \triangleq \blkrow\{\bm \Phi^{(1,n)}[k],\ldots,\bm \Phi^{(M_\text{tck},n)}[k]\}$, and $\bsfn^{(n)}[k] = [\bsfn^{(1,n)T}[k], \ldots, \bsfn^{(M_\text{tck},n)T}[k]]^T$ allows us to extend \eqref{equation:vect_signal_model_single_tck} for $M_\text{tck}$ received training OFDM symbols as

\begin{equation}
\bsfy^{(n)}[k] \approx \bm \Phi^{(n)}[k] \vect\{\bsfH^{(n)}[k]\} + \bsfn^{(n)}[k],
	\label{equation:non_linear_CS_model}
\end{equation}
where $\vect\{\bsfH^{(n)}[k]\}$ can be written as 
\begin{equation}
\begin{split}
	\vect\{\bsfH^{(n)}[k]\} \approx& \underbrace{\left(\bA_\text{T}^\text{C}\left(\bm \theta^{(n)}\right) \circ \bA_\text{R}\left(\bm \phi^{(n)}\right)\right)}_{\bm \Psi\left(\bm \theta^{(n)},\bm \phi^{(n)}\right)} \\
	&\times \underbrace{\vect\{\diag\{\bsfG^{(n)}[k]\}\}}_{\bsfg^{(n)}[k]}.
\end{split}
\end{equation}
Therefore, \eqref{equation:non_linear_CS_model} can be expressed in terms of the unknown AoA/AoD and channel gains as
\begin{equation}
	\bsfy^{(n)}[k] \approx \bm \Phi^{(n)}[k] \bm \Psi\left(\bm \theta^{(n)},\bm \phi^{(n)}\right) \bsfg^{(n)}[k] + \bsfn^{(n)}[k].
	\label{equation:non_linear_CS_complete}
\end{equation}

The signal model in \eqref{equation:non_linear_CS_complete} can be used to derive different channel tracking algorithms. Nonetheless, using frequency-dependent training symbols $\bsfs^{(m,n)}[k]$ leads to more computationally complex algorithms. Then, to simplify subsequent calculations, we may express the training streams during tracking $\bsfs^{(m,n)}[k]$ as $\bsfs^{(m,n)}[k] = \bsfq^{(m,n)} \sfs^{(m,n)}[k]$, where $\bsfq^{(m,n)} \in \mathbb{C}^{\Lt \times 1}$ is a frequency-flat spatial modulation vector consisting of independent and identically distributed energy-normalized QPSK constellation symbols, and $\sfs^{(m,n)}[k]$ is a frequency-dependent constellation symbol (i.e. QPSK) whose effect can be eliminated at the receiver by simply multiplying $\bsfy^{(n)}[k]\sfs^{(m,n)-1}[k]$, without altering noise statistics. Further, as shown in \cite{RodGonVenHea:TWC_18}, the use of $\bsfq^{(m,n)}$ allows exploiting the $\Lt$ degrees of freedom coming from the transmit hybrid \ac{MIMO} architecture. Thereby, \eqref{equation:non_linear_CS_complete} can be simplified to
\begin{equation}
	\bsfy^{(n)}[k] \approx \bm \Phi^{(n)} \bm \Psi\left(\bm \theta^{(n)},\bm \phi^{(n)}\right) \bsfg^{(n)}[k] + \bsfn^{(n)}[k].
	\label{equation:non_linear_CS_complete_flat}
\end{equation}

In the following section, we propose two novel algorithms to track variations of \ac{mmWave} frequency-selective \ac{MIMO} channels. The first algorithm aims at tracking off-grid variations of \ac{AoA} and \ac{AoD} following the \ac{ML} philosophy, i.e., trying to maximize the \ac{LLF} of the received data without relying on the extended virtual channel model \cite{HeaPreRanRohSay:Overview-Sig-Proc-mmWaveMIMO:16}. The second algorithm, however, aims at tracking both the off-grid variations of \ac{AoA}, \ac{AoD}, and channel gains in the frequency domain, by exploiting prior statistical knowledge on the \ac{SSP} parameters. 

\subsection{Subcarrier Selection - Simultaneous Iterative Gridless Weighted - Orthogonal Least Squares (SS-SIGW-OLS)}\label{sec:Gradient_Tracking}

In this subsection, our first proposed channel tracking algorithm is introduced. Its underlying principle is to track variations of \ac{AoA} and \ac{AoD} following the \ac{ML} criterion. Thereafter, the channel gains are obtained using the \ac{MVUE}, which has been shown to be optimal in \cite{RodGonVenHea:TWC_18}. Finally, the channel matrices are reconstructed using the estimated \ac{AoA}, \ac{AoD}, channel gains, and knowledge on the array geometry.

We denote the estimate of the sparsity level as $\hat{L}$, and $\hat{\bm \theta}^{(n-1)}$, $\hat{\bm \phi}^{(n-1)} \in \mathbb{R}^{\hat{L} \times 1}$ are the estimates of the AoD and AoA obtained for the $(n-1)$-th channel slot. Let us define an $M\Lr K \times 1$ vector containing the received wideband signal as $\bsfy^{(n)} \triangleq \left[\begin{array}{ccc} \bsfy^{(n)T}[0] & \ldots & \bsfy^{(n)T}[K-1] \\ \end{array}\right]^T$, and a $K\hat{L} \times 1$ vector $\bsfg^{(n)}$ containing the complex channel gains for the entire communications band as $\bsfg^{(n)} \triangleq \left[\begin{array}{ccc} \bsfg^{(n)T}[0] & \ldots \bsfg^{(n)T}[K-1] \\ \end{array}\right]^T$. Let us also express $\bC_\text{w}^{(m,n)} \triangleq \bD_\text{w}^{(m,n)*} \bD_\text{w}^{(m,n)}$, with $\bD_\text{w}^{(m,n)} \in \mathbb{C}^{M\Lr \times M\Lr}$ the Cholesky factor of $\bC_\text{w}^{(m,n)}$. Likewise, we define $\bD_\text{w}^{(n)} = \blkdiag\{\bD_\text{w}^{(1,n)},\ldots,\bD_\text{w}^{(\Mtck,n)}\}$. Now, let $\bsfy_\text{w}^{(n)}[k] \in \mathbb{C}^{\Mtck \Lr \times 1}$ be the post-whitened received signal at $k$-th subcarrier, and $\bm \Upsilon_\text{w}^{(n)} \in \mathbb{C}^{\Mtck \Lr \times \hat{L}}$ be the post-whitened equivalent measurement matrix. These matrices are defined in a similar manner as in \cite{RodGonVenHea:TWC_18}:
\begin{equation}
\begin{split}
	\bsfy_\text{w}^{(n)}[k] &\triangleq \bsfD_\text{w}^{(n)-*} \bsfy^{(n)}[k] \\
	\bm \Upsilon_\text{w}^{(n)}\left(\bm \theta^{(n)},\bm \phi^{(n)}\right) &\triangleq \underbrace{\bsfD_\text{w}^{(n)-*} \bm \Phi^{(n)}}_{\bm \Phi_\text{w}^{(n)}} \bm \Psi\left(\bm \theta^{(n)},\bm \phi^{(n)}\right).
\end{split}
\label{equation:post_whitened_matrices}
\end{equation} 
Using \eqref{equation:post_whitened_matrices}, we can define the projection matrix onto the subspace spanned by $\bm \Upsilon_\text{w}^{(n)}$ as $\bP\left(\bm \theta^{(n)},\bm \phi^{(n)}\right) \in \mathbb{C}^{\Mtck \Lr \times \Mtck \Lr}$, which is given by
\begin{equation}
	\bP\left(\bm \theta^{(n)},\bm \phi^{(n)}\right) \triangleq \bm \Upsilon_\text{w}^{(n)}\left(\bm \theta^{(n)},\bm \phi^{(n)}\right)\bm \Upsilon_\text{w}^{(n)\dag}\left(\bm \theta^{(n)},\bm \phi^{(n)}\right).
	\label{equation:proj_matrix_Upsilon}
\end{equation}
The insight behind this equation is as follows: according to \eqref{equation:non_linear_CS_model}, the received signal is a point in $\mathbb{C}^{\Mtck \Lr}$ consisting of a compressed linear combination of the vectorized channel coefficients. The vectorized channel is a linear combination of a function of the array matrices evaluated in the true \ac{AoA} and \ac{AoD}. The projection matrix in \eqref{equation:proj_matrix_Upsilon} allows us to measure the extent to which $\bsfy^{(n)}[k]$ lies within the column space of the vectorized channel, and it is thus the key to properly estimate the \ac{AoA}, \ac{AoD}, and channel gains, as we will show next. 

Remark: the cost function in \eqref{equation:proj_matrix_Upsilon} is a projection function, which allows finding the \ac{AoA}/\ac{AoD} using successive iterations \cite{Globecom_2018_Beam_Squint}. Using a correlation function as in \cite{RodGonVenHea:TWC_18}, is, however, suboptimal, and it only exhibits the same asymptotic performance as the function in \eqref{equation:proj_matrix_Upsilon} when $K$ and $M_\text{tck}$ are large enough \cite{Globecom_2018_Beam_Squint}. Therefore, we choose to maximize \eqref{equation:proj_matrix_Upsilon} to exploit the couplings between the different angular parameters to obtain the best possible performance, as shown in \cite{Globecom_2018_Beam_Squint}.

 The \ac{LLF} of $\bsfy^{(n)}$ has been shown to fulfil \cite{CAMSAP_2017_Offgrid}
\begin{equation}
	\ln p \left(\bsfy^{(n)}; \bm \theta^{(n)},\bm \phi^{(n)}, \bsfg^{(n)}\right) \propto \sum_{k=0}^{K-1}{\bsfy_\text{w}^*[k] \bP\left(\bm \theta^{(n)},\bm \phi^{(n)}\right) \bsfy_\text{w}[k]},
	\label{equation:llf_propto}
\end{equation}
such that the problem of finding the ML estimator for the AoA and AoD can be stated as
\begin{equation}
	\left\{\hat{\bm \theta}_\text{ML}^{(n)}, \hat{\bm \phi}_\text{ML}^{(n)}\right\} = \underset{\bm \theta^{(n)},\bm \phi^{(n)}}{\arg\,\max}\,\sum_{k=0}^{K-1}{\underbrace{\bsfy_\text{w}^{(n)}[k]\bP\left(\bm \theta^{(n)},\bm \phi^{(n)}\right) \bsfy_\text{w}^{(n)}[k]}_{{\cal L}_k\left( \bm \theta^{(n)},\bm \phi^{(n)} \right)}},
	\label{equation:optimization_gridless}
\end{equation}
Although the function in \eqref{equation:optimization_gridless} is non convex due to the non linear dependence on $\bm \theta^{(n)}$, $\bm \phi^{(n)}$, it can be maximized using an iterative gradient ascent approach. To guarantee convergence of the gradient approach, it is necessary that the function to optimize is locally convex in a neighborhood around the initial point. It can be shown that the function is indeed locally convex since the \ac{SS-SW-OMP+Th} converges to a local optimum if the dictionary sizes are large enough \cite{RodGonVenHea:TWC_18}. Then, a suitable initial point is given by the estimates of the \ac{AoA}s and \ac{AoD}s found at the $(n-1)$-th channel slot, since they are locally optimum if the number of measurements and/or subcarrieres is large enough \cite{RodGonVenHea:TWC_18}. We can, however, reduce computational complexity by processing a smaller number of subcarriers. Let ${\cal K}_\text{p}$ be the set of indexes of the $K_\text{p}$ subcarriers that exhibit the largest $\ell_2$-norm (equivalently, the largest received $\SNR$), similar to \cite{RodGonVenHea:TWC_18}. Then, if we define a stepsize matrix $\bm \Lambda \in \mathbb{R}_{+}^{2\hat{L} \times 2\hat{L}}$, the update equation for the \ac{AoA} and \ac{AoD} is given by
\begin{equation}
\begin{split}
	\left[\begin{array}{c} \hat{\bm \theta}_\text{ML}^{(n,i)} \\ \hat{\bm \phi}_\text{ML}^{(n,i)} \\ \end{array}\right] &= \underbrace{\left[\begin{array}{c} \hat{\bm \theta}_\text{ML}^{(n,i-1)} \\ \hat{\bm \phi}_\text{ML}^{(n,i-1)} \\ \end{array}\right]}_{\text{Current guess}} \\ &+ \underbrace{\bm \Lambda \sum_{k \in {\cal K}_\text{p}} \left[\begin{array}{c} \bm \nabla_{\bm \theta}{\cal L}_k\left(\bm \theta^{(n)},\bm \phi^{(n)}\right)\rvert_{\bm \theta^{(n)} = \hat{\bm \theta}_\text{ML}^{(n,i-1)}} \\ 
	\bm \nabla_{\bm \phi}{\cal L}_k\left(\bm \theta^{(n)},\bm \phi^{(n)}\right)\rvert_{\bm \phi^{(n)} = \hat{\bm \phi}_\text{ML}^{(n,i-1)}} \\ \end{array}\right]}_{\text{Correction factor}},
\end{split}
	\label{equation:gradient_ML}
\end{equation}
in which $\hat{\bm \theta}_\text{ML}^{(n,0)}$, $\hat{\bm \phi}_\text{ML}^{(n,0)}$ are initialized to the estimates of $\bm \theta$, $\bm \phi$ found during the $(n-1)$-th channel slot. The form of the gradient vectors $\bm \nabla_{\bm \theta}\ln{p\left(\bsfy;\bm \theta^{(n)},\bm \phi^{(n)}\right)}$ and $\bm \nabla_{\bm \phi}\ln{p\left(\bsfy;\bm \theta^{(n)},\bm \phi^{(n)}\right)}$ is derived in Appendix~A. After a sufficient number of iterations $I$ to allow convergence, the estimated angles are expected to be sufficiently close to the true angles of the mmWave \ac{MIMO} channel. Thereafter, the channel gains are estimated using the Minimum Variance Unbiased Estimator (MVUE), which is optimal as shown in \cite{RodGonVenHea:TWC_18}
\begin{equation}
	\hat{\bsfg}^{(n)}[k] = \left[\bm \Upsilon_\text{w}\left(\hat{\bm \theta}_\text{ML}^{(I)},\hat{\bm \phi}_\text{ML}^{(I)}\right)\right]^\dag \bsfy_\text{w}^{(n)}[k].
	\label{equation:channel_gains_MVUE}
\end{equation}
Finally, the channel matrices $\bsfH^{(n)}[k]$, $k = 0,\ldots,K-1$ are reconstructed using the estimates for the AoA, AoD, and channel gains as
\begin{equation}
	\widehat{\bsfH}^{(n)}[k] = \vect^{-1}\left\{ \left(\hat{\bA}_\text{T}^{\text{C}}\left(\hat{\bm \theta}_\text{ML}^{(n,I)}\right) \circ \hat{\bA}_\text{R}\left(\hat{\bm \phi}_\text{ML}^{(n,I)}\right)\right) \hat{\bsfg}^{(n)}[k] \right\},
\end{equation}
where the operator $\vect^{-1}\{\cdot \}$ simply reshapes the input vector into matrix form. The detailed steps of the proposed \ac{SS-SIGW-OLS} algorithm are summarized in Algorithm \ref{alg:SSSIGWOLS}.

\begin{algorithm}
\caption{Subcarrier Selection - Simultaneous Iterative Gridless Weighted - Orthogonal Least Squares (OLS)}\label{alg:SSSIGWOLS}
\begin{algorithmic}[1]
\Procedure{SS-SIGW-OLS($\bsfy_\text{w}[k]$,$\bm \Phi_\text{w}$,$\bm \Lambda$,$\hat{L}$,$\hat{\bm\theta}^{(n-1)},\hat{\bm\phi}^{(n-1)}$)}{}
\State \textbf{Initialize counter and residual vectors} \\\vspace{0.1cm}
\qquad $s = 0$,\qquad $i = 0$,\quad $\bsfr[k] = \bsfy_\text{w}[k]$, $k = 0,\ldots,K-1$ \vspace{0.15cm}
\State \textbf{Find the $K_\text{p}$ strongest subcarriers}
\qquad \While {$i \leq K_\text{p}$} \\
\qquad \qquad ${\cal K} = {\cal K} \cup \underset{k\not\in {\cal K}}{\arg\,\max}\,\|\bsfy_\text{w}[k]\|_2^2$ \\
\qquad \qquad $i = i + 1$ 
\EndWhile \\
 \qquad \qquad $\hat{\bm \phi}_\text{ML}^{(n,0)} = \hat{\bm \phi}_\text{ML}^{(n-1)}$ \\
 \qquad \qquad $\hat{\bm \theta}_\text{ML}^{(n,0)} = \hat{\bm \theta}_\text{ML}^{(n-1)}$ \\
\State \textbf{Maximum-Likelihood optimization}\\
\qquad \For {$i=1:I$} \\\\
\qquad $\bsfee^{(n,i)} = \left[\begin{array}{c} \bm \nabla_{\bm \theta}{\cal L}_k\left(\bm \theta^{(n)},\bm \phi^{(n)}\right)\rvert_{\bm \theta^{(n)} = \hat{\bm \theta}_\text{ML}^{(n,i-1)}} \\ 
	\bm \nabla_{\bm \phi}{\cal L}_k\left(\bm \theta^{(n)},\bm \phi^{(n)}\right)\rvert_{\bm \phi^{(n)} = \hat{\bm \phi}_\text{ML}^{(n,i-1)}} \\ \end{array}\right]$ \\
\qquad  $\left[\begin{array}{c}
\bm \theta_\text{ML}^{(n,i)} \\
\bm \phi_\text{ML}^{(n,i)} \\ \end{array}\right] = \left[\begin{array}{c}
\bm \theta_\text{ML}^{(n,i-1)} \\
\bm \phi_\text{ML}^{(n,i-1)} \\ \end{array}\right] + \bm \Lambda \sum_{k\in {\cal K}_\text{p}}\bsfee^{(n,i)}$
\qquad \qquad \qquad \EndFor
\qquad \State \textbf{Estimate channel gains according to the estimated angles} \\ \vspace{0.1cm}
\qquad \qquad $\hat{\bm \Psi}_\text{ML} = \bA_\text{T}^\text{C}\left(\hat{\bm \theta}_\text{ML}^{(n,I)}\right) \circ \bA_\text{R}\left(\hat{\bm \phi}_\text{ML}^{(n,I)}\right)$ \\
\qquad \qquad $\hat{\bm \Upsilon}_\text{w} = \bm \Phi_\text{w} \hat{\bm \Psi}_\text{ML}$ \\
\qquad \qquad $\hat{\bsfg}^{(n)}_{\text{ML}}[k] = \hat{\bm \Upsilon}_\text{w}^\dag\bsfy_\text{w}[k]$, $k = 0,\ldots,K-1$ \\
\qquad \qquad $\hat{\bsfH}_\text{ML}^{(n)}[k] = \unvect\left\{\hat{\bm \Psi}_\text{ML} \hat{\bsfg}_\text{ML}^{(n)}[k]\right\}$ \\
\EndProcedure
\end{algorithmic}
\end{algorithm}

\subsection{Generalized Marginalized Particle Filter}

In this subsection, we present an alternative channel tracking algorithm. Unlike the \ac{SS-SIGW-OLS} approach proposed in the previous section, this second strategy exploits prior statistical knowledge about the \ac{AoA}, \ac{AoD}, and channel gains. 

To track off-grid variations of the \ac{AoA} and \ac{AoD}, some prior knowledge on the sparsity level (number of significant multipath components) is required. Channel models at \ac{mmWave} are usually defined in terms of average number of clusters and average number of rays per cluster \cite{5G_channel_model}, \cite{QuaDRiGa_IEEE}, \cite{QuaDRiGa_Tech_Rep}, \cite{NYUSIM}, for which there is prior statistical knowledge. Therefore, to devise our Bayesian tracking strategy, we consider that, on average, there are $\mu_L = C\mu_{R_\text{c}}$ multipath components in the channel, with $\mu_{R_\text{c}}$ the average number of rays per cluster and $C$ the average number of clusters.

\begin{figure}[ht!]
\centering
\includegraphics[width=0.7\columnwidth]{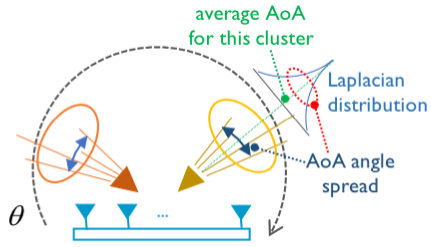}     
    \caption{Illustration of the prior statistical knowledge about the he conditional density function $p(\bm \xi^{(n)}|\bm \xi^{(n-1)})$.}
    \label{fig:statisticsAoA}
\end{figure}

We also consider that there is prior  information about the distributions of the \ac{AoA} and \ac{AoD}. We define $\bm \xi^{(n)}=[\bm \theta^{(n)T}, \bm \phi^{(n)T}]^T, \in \mathbb{R}^{2\mu_L \times 1}$ as the set of \ac{AoA} and \ac{AoD} for the $n$-th channel slot, and $\bm \xi^{(n-1)}$ the set of \ac{AoA} and \ac{AoD} for the $(n-1)$-th channel slot. The density of $\bm \xi^{(n)}$ can be arbitrary, but it is assumed known. In this paper, we assume, as in the 5G NR channel model, that the conditional density function $p(\bm \xi^{(n)}|\bm \xi^{(n-1)})$ follows a Laplacian distribution for each multipath component. We compute the mean of this Laplacian distribution from the channel estimate in slot 0, grouping the rays within clusters. This way, the conditional distribution for each ray is Laplacian with mean the sample mean of the cluster obtained from the initial channel estimation, as illustrated in Fig.~\ref{fig:statisticsAoA}. The variance is computed as the square of the RMS angular spread for AoA/AoD provided by the selected channel model (see for example Table 7.5-6, Part 1 in \cite{5G_channel_model}). 

Likewise, let $\bsfg^{(n)} \in \mathbb{C}^{K\mu_L \times 1}$ denote the set of channel gains for the different subcarriers, $\bsfg^{(n)} = [\bsfg^{(n)T}[0],\ldots,\bsfg^{(n)T}[K-1]]^T$. Similarly to \cite{ICC_2019_ERBPF}, we assume that that this vector follows a first-order Gauss-Markov process 
\begin{equation}
	\bsfg^{(n)} = \bsfR^{(n)} \bsfg^{(n-1)} + \bsfX^{1/2}\bm \Delta \bsfg^{(n)},
	\label{equation:evolution_channel_gains}
\end{equation}
whereby the conditional pdf of $\bsfg^{(n)}|\bsfg^{(n-1)}$ follows $\bsfg^{(n)}|\bsfg^{(n-1)} \sim {\cal CN}\left(\bsfR^{(n)} \bsfg^{(n-1)},\bsfX^{(n)}\right)$, for certain $\bsfR^{(n)}$, $\bsfX^{(n)}$, which are assumed to be known. Further, $\bsfg^{(n)} \sim {\cal CN}\left(\bm 0, \bsfC_{\bsfg\bsfg}^{(n)}\right)$, with $\bsfC_{\bsfg\bsfg}^{(n)} = \bsfR^{(n)} \bsfC_{\bsfg\bsfg}^{(n-1)} \bsfR^{(n)*} + \bsfX^{(n)}$. In the numerical results section, we will particularize the choices of $\bsfC_{\bsfg\bsfg}^{(n)}$, $\bsfR^{(n)}$, $\bsfX^{(n)}$, as well as justify how to obtain them in practice. 

In the following, we will derive our Bayesian filtering strategy based on these assumptions. In particular, we will focus on finding the \textit{sparsity-constrained} \ac{MMSE} estimator of $\{\bsfH^{(n)}[k]\}_{k=0}^{K-1}$. Let $\bsfy^{(n)} = [\bsfy^{(n)T}[0], \ldots, \bsfy^{(n)T}[K-1]]^T$ denote the received wideband signal, and $\bsfT\left(\bm \xi^{(n)}\right) = \bI_K \otimes \bm \Phi^{(n)} \bm \Psi\left(\bm \xi^{(n)}\right)$ denote the transfer matrix of the system. Therefore, the wideband received signal yields
\begin{equation}
	\bsfy^{(n)} \approx \left(\bI_K \otimes \bm \Phi^{(n)} \bm \Psi\left(\bm \xi^{(n)}\right)\right) \bsfg^{(n)} + \bsfn^{(n)}.
	\label{equation:vectorized_rx_signal_full_wideband}
\end{equation}
Let us define $\bsfC_\text{w}^{(n)} = \blkdiag\{\bsfC_\text{w}^{(1,n)},\ldots,\bsfC_\text{w}^{(M_\text{tck},n)}\}$. Then, $\bsfn^{(n)}$ in \eqref{equation:vectorized_rx_signal_full_wideband} follows $\bsfn^{(n)} \sim {\cal CN}\left(\bm 0, \sigma^2 \bI_K \otimes \bsfC_\text{w}^{(n)}\right)$. The measurement model in \eqref{equation:vectorized_rx_signal_full_wideband} is non-linear in $\bm \xi^{(n)}$, but the received signal $\bsfy^{(n)}$ exhibits a linear sub-structure when conditioned on $\bm \xi^{(n)}$, i.e., $\bsfy^{(n)}|\bm \xi^{(n)} \sim {\cal CN}\left(\bsfT\left(\bm \xi^{(n)}\right)\bsfC_{gg}^{(n)}\bsfT^*\left(\bm \xi^{(n)}\right) + \bsfC_\text{w}^{(n)}\right)$. Therefore, exploiting this signal structure can be used to obtain better estimates by analytically marginalizing out the linear state variables $\bsfg^{(n)}$. The problem of finding the \ac{MMSE} estimator of $\bsfh^{(n)} = \vect\{[\bsfH^{(n)}[0],\ldots,\bsfH^{(n)}[K-1]]\} = \left(\bI_K \otimes \bm \Psi\left(\bm \xi^{(n)}\right)\right)\bsfg^{(n)}$ reduces to finding its posterior conditional expectation \cite{Kay:Fundamentals-of-Statistical-Signal:93} as ($\bm \Xi^{(n)} = \{\bm \xi^{(i)}\}_{i=0}^{n}$, $\bsfY^{(n)} = \{\bsfy^{(i)}\}_{i=0}^{n}$)
\begin{equation}
	\hat{\bsfh}_\text{MMSE}^{(n)} \triangleq \mathbb{E}_{\bsfh^{(n)}|\bsfY^{(n)}}\{\bsfh^{(n)}\}.
	\label{equation:MMSE_problem}
\end{equation}
As shown in \eqref{equation:joint_posterior_pdf}, an approximation of the \ac{MMSE} estimator in \eqref{equation:MMSE_problem} will be given by the combination of a \ac{KF} to estimate the channel gains $\bsfg^{(n)}$, a \ac{PF} to estimate the angular parameters in $\bm \xi^{(n)}$, and a second \ac{PF} to estimate the vectorized wideband channel $\bsfh^{(n)}$.
The joint posterior pdf of $\bsfg^{(n)}$, $\bm \Xi^{(n)}$, and $\bsfh^{(n)}$ is given by
\begin{equation}
\begin{split}
	p(\bsfg^{(n)},\bm \Xi^{(n)},\bsfh^{(n)}|\bsfY^{(n)}) = & \underbrace{p(\bsfh^{(n)}|\bsfY^{(n)},\bsfg^{(n)},\bm \Xi^{(n)})}_{\text{PF \# 2}} \\
	& \times \underbrace{p(\bm \Xi^{(n)}|\bsfY^{(n)})}_{\text{PF \# 1}} \underbrace{p(\bsfg^{(n)}|\bm \Xi^{(n)},\bsfY^{(n)})}_{\text{Optimum KF}}.
\end{split}
\label{equation:joint_posterior_pdf}
\end{equation}
The different terms in the right hand side of \eqref{equation:joint_posterior_pdf} can be seen to yield
\begin{equation}
	p(\bsfh^{(n)} | \bsfY^{(n)}, \bsfg^{(n)}, \bm \Xi^{(n)}) = \delta\left(\bsfh^{(n)} - \left(\bI_K \otimes \bm \Psi\left(\bm \xi^{(n)}\right)\right) \bsfg^{(n)}\right)
	\label{equation:pdf_EPF}
\end{equation}
\begin{equation}
\begin{split}
	p(\bm \Xi^{(n)} | \bsfY^{(n)}) &\propto p(\bsfy^{(n)} | \bm \Xi^{(n)}, \bsfY^{(n-1)}) \times \\ &\times p(\bm \xi^{(n)} | \bm \Xi^{(n-1)}, \bsfY^{(n-1)}) \times \\ &\times p (\bm \Xi^{(n-1)} | \bsfY^{(n-1)})
\end{split}
	\label{equation:pdf_PF}
\end{equation}
\begin{equation}
	p(\bsfg^{(n)}| \bm \Xi^{(n)}, \bsfY^{(n)}) = {\cal CN}\left(\hat{\bsfg}^{(n|n)}, \bsfC_{\bsfg\bsfg}^{(n|n)}\right),
\label{equation:pdf_KF}
\end{equation}
\begin{equation}
	p(\bsfg^{(n|n-1)}|\bm \Xi^{(n)}, \bsfY^{(n)}) = {\cal CN}\left(\hat{\bsfg}^{(n|n-1)}, \bsfC_{\bsfg\bsfg}^{(n|n-1)}\right),
	\label{equation:pdf_prediction_KF}
\end{equation}
wherein $p(\bx) = {\cal CN}(\bm \mu, \bC)$ indicates that $p(\bx)$ takes the form of a circularly symmetric multivariate complex Gaussian distribution with mean $\bm \mu$ and covariance matrix $\bC$. The last two pdfs in \eqref{equation:pdf_KF} and \eqref{equation:pdf_prediction_KF} are of the form of the optimal \ac{KF}, whereby exploiting the linear sub-structure in \eqref{equation:vectorized_rx_signal_full_wideband} is clearly emphasized. Using straightforward application of the \ac{KF} \cite{Kay:Fundamentals-of-Statistical-Signal:93}, \cite{Kalman_Filter_1960}, the statistics in \eqref{equation:pdf_KF}-\eqref{equation:pdf_prediction_KF} yield the well-known \ac{KF} prediction and update equations \cite{MPF_IEEE},\cite{Kay:Fundamentals-of-Statistical-Signal:93}, which are omitted here for the sake of brevity.

The second pdf in \eqref{equation:pdf_KF} does not admit a closed-form for the different terms on its right hand side. Nonetheless, the terms $p\left(\bsfy^{(n)}|\bm \Xi^{(n)},\bsfY^{(n-1)}\right)$, $p\left(\bm \xi^{(n)}|\bm \Xi^{(n-1)},\bsfY^{(n-1)}\right)$ can be expressed in closed form. Thus, we will approximate it using a standard \ac{PF}, which corresponds to the \ac{PF} $\# 1$. By direct inspection of $\bsfy^{(n)}$ in \eqref{equation:vectorized_rx_signal_full_wideband}, assuming that the noise realizations for two different channel slots are independent, it follows that the random vector $\bsfz^{(n)} = \bsfy^{(n)}|\bm \Xi^{(n)},\bsfY^{(n-1)}$ is distributed as $\bsfz^{(n)} \sim {\cal CN}\left(\bm \mu_{\bsfz}^{(n)}, \bsfC_{\bsfz\bsfz}^{(n)}\right)$, with $\bm \mu_z^{(n)}$, $\bsfC_z^{(n)}$ given by
\begin{equation}
\begin{split}
 \bm \mu_z^{(n)} &= \bsfT\left(\bm \xi^{(n)}\right)\hat{\bsfg}^{(n|n-1)} \\
 \bsfC_{\bsfz\bsfz}^{(n)} &= \bsfT\left(\bm \xi^{(n)}\right) \bsfC_{\bsfg\bsfg}^{(n|n-1)} \bsfT^*\left(\bm \xi^{(n)}\right) + \sigma^2\bsfC_\text{w}.
\end{split}
\end{equation}
The other random vector $\bm \xi^{(n)}|\bm \Xi^{(n-1)},\bsfY^{(n-1)}$ is distributed with pdf $p(\bm \xi^{(n)}|\bm \xi^{(n-1)})$, which is assumed to be known, yet arbitrary. The pdf of $\bm \Xi^{(n-1)}|\bsfY^{(n-1)}$ corresponding to the last term in \eqref{equation:pdf_PF} can be approximated by the previous iteration of the first \ac{PF} highlighted in \eqref{equation:joint_posterior_pdf}. Now, the linear system in \eqref{equation:evolution_channel_gains} and \eqref{equation:vectorized_rx_signal_full_wideband} can be formed for the $i$-th particle $\{\bm \xi^{(n,i)}\}_{i=1}^{N_\text{PF}}$, $1 \leq i \leq N_\text{PF}$, in which $N_\text{PF}$ is the number of particles used by the \ac{PF}. This requires one \ac{KF} associated with each particle. Let the final sampled distribution of $\bm \Xi^{(n)}|\bsfY^{(n)}$ be denoted by the weights $\left\{w\left(\bm \Xi^{(n,i)}\right)\right\}_{i=1}^{N_\text{PF}}$, which must satisfy $\sum_{i=1}^{N_\text{PF}}w\left(\bm \Xi^{(n,i)}\right) = 1$, since we are approximating a pdf by a probability mass function (pmf). Let us define $\bsfB\left(\bm \xi^{(n)}\right) = \bI_K \otimes \bm \Psi\left(\bm \xi^{(n)}\right)$. Then, the vectorized channel can be estimated as
\begin{equation}
\begin{split}
	\hat{\bsfh}^{(n)} &= \int_{{\cal S}\left(\bsfg^{(n)},\bm \Xi^{(n)}\right)} \bsfB\left(\bm \xi^{(n)}\right) \bsfg^{(n)} \\ &\times \underbrace{p(\bsfg^{(n)}|\bm \Xi^{(n)},\bsfY^{(n)}) w\left(\bm \Xi^{(n,i)}\right)}_{q^{(n)}} d\bsfg^{(n)} d\bm \Xi^{(n)}.
\end{split}
	\label{equation:MMSE_estimator_channel}
\end{equation}
The final estimator in \eqref{equation:MMSE_estimator_channel} involves a complex integration over the joint support of $\bsfg^{(n)},\bm \Xi^{(n)}$, which cannot be solved in closed form. For this reason, we propose to approximate it using a second \ac{PF}, which was highlighted in \eqref{equation:joint_posterior_pdf}. Thus, normalizing the weights $q^{(n)}$ so that $\tilde{q}^{(n,i)} = q^{(n,i)}/\sum_{n=1}^{N_\text{PF}}q^{(n,i)} \tilde{q}^{(n-1,i)}$, the \ac{MMSE} estimator of the channel can be approximately expressed as a convex combination of $\bsfB\left(\bm \xi^{(n,i)}\right)$ and $\hat{\bsfg}^{(n|n,i)}$ 
\begin{equation}
	\hat{\bsfh}^{(n)} \approx \sum_{n=1}^{N_\text{PF}}\tilde{q}^{(n,i)} \bsfB\left(\bm \xi^{(n,i)}\right) \hat{\bsfg}^{(n|n,j)}.
	\label{equation:MMSE_estimator_approx}
\end{equation}
\textbf{Remark}: for the \ac{ML} estimator, a function $f(\bsfz)$ of \ac{ML} estimates $\hat{\bsfz}_\text{ML}$ will generally yield the \ac{ML} estimator of a given vector $\bsfz$ \cite{Kay:Fundamentals-of-Statistical-Signal:93}. In the Bayesian framework, this property, called the \textit{asymptotic invariance property} \cite{Kay:Fundamentals-of-Statistical-Signal:93} does not hold. We might be tempted to directly calculate the estimator of the channel by using the \ac{MMSE} estimator of $\bm \xi^{(n)}$ and $\bsfg^{(n)}$, but this would not result in the \ac{MMSE} estimator of $\bsfh^{(n)}$. To find the true \ac{MMSE} estimator, the pdf of $\bsfh^{(n)} = f\left(\bm \xi^{(n)},\bsfg^{(n)}\right)$ needs to be calculated to find the \textit{posterior conditional mean}, as shown in \eqref{equation:MMSE_estimator_channel}.

In general, for a finite number of particles $N_\text{PF}$, there will be an approximation error in \eqref{equation:MMSE_estimator_approx}. The details on how large this approximation error is are beyond the scope of this work, and will thus be devoted to a future manuscript. It is important, however, to highlight that our proposed approach is a generalization of the \ac{MPF} presented in \cite{MPF_IEEE}, which has already been shown to perform well in practice and has optimality guarantees as $N_\text{PF}$ grows large. Therefore, our proposed algorithm can be also expected to work well in practice, as it will be shown in numerical results. Owing to our proposed algorithm being based on the \ac{MPF} in \cite{MPF_IEEE}, we name the proposed algorithm as \ac{GMPF}. The detailed steps the  algorithm follows are given in Algorithm \ref{alg:GMPF}.

\begin{algorithm}
\caption{Generalized Marginalized Particle Filter (GMPF)}\label{alg:GMPF}
\begin{algorithmic}[1]
\State{GMPF${\big{(}}\bsfY^{(n)}$,$N_\text{PF}$,$\bm{\Phi}^{(n)},\hat{\theta}_{r,u}^{\text{azi}(0)}$,$\hat{\theta}_{r,u}^{\text{ele}(0)}$,$\hat{\phi}_{r,u}^{\text{azi}(0)}$,$\hat{\phi}_{r,u}^{\text{ele}(0)}{\big{)}}$}
\State \textbf{Initialize angular particles based on prior information} \\
\qquad $\theta_{r,u}^{\text{azi}(1|0,i)} = \hat{\theta}_{r,u}^{\text{azi}(0)} + {\cal L}(0,\sigma_{\Delta,u}^{\text{azi}2})$\\
\qquad $\theta_{r,u}^{\text{ele}(1|0,i)} = \hat{\theta}_{r,u}^{\text{ele}(0)} + {\cal L}(0,\sigma_{\Delta,u}^{\text{ele}2})$\\
\qquad $\phi_{r,u}^{\text{azi}(1|0,i)} = \hat{\phi}_{r,u}^{\text{azi}(0)} + {\cal L}(0,\sigma_{\Delta,u}^{\text{azi}2})$\\
\qquad $\phi_{r,u}^{\text{ele}(1|0,i)} = \hat{\phi}_{r,u}^{\text{ele}(0)} + {\cal L}(0,\sigma_{\Delta,u}^{\text{ele}2})$\\
\State \textbf{Initialize linear particles} \\
\qquad $\hat{\bsfg}^{(1|0,i)} =  \bm 0$, \quad $\hat{\bsfC}_{\bsfg\bsfg}^{(1|0,i)} = \frac{\Nt \Nr}{\mu_L} \bI_{\mu_L} \otimes \bsfF_1 \bsfF_1^*$, with $\bsfF_1 \in \mathbb{C}^{K \times Z_\text{p}}$ the matrix comprising the first $Z_\text{p}$ Fourier vectors.
\State \textbf{Evaluate the importance weights $q^{(n,i)} = p(\bsfy^{(n)}|\bm \Xi^{(n,i)},\bsfY^{(n-1)})p(\bsfg^{(n)}|\bm \Xi^{(n)},\bsfY^{(n)})$ and normalize them}\\
\qquad $\tilde{q}^{(n,i)} = \frac{q^{(n,i)}}{\sum_{j=1}^{N_\text{PF}}q^{(n,j)}} \tilde{q}^{(n-1,i)}$\\
\State \textbf{PF} $\#1$ \textbf{measurement update (resampling} $N_\text{PF}$ \textbf{particles with replacement)} \\
\qquad $\mathbb{P}\left(\bm \xi^{(n|n,i)} = \bm \xi^{(n|n-1,j)}\right) = \tilde{q}^{(n,j)}$\\
\State \textbf{KF measurement update for $1\leq i \leq N_\text{PF}$} \\
\qquad $\bsfK^{(n,i)} = \bsfC_{\bsfg\bsfy}^{(n|n-1,i)}\bsfC_{\bsfy\bsfy}^{(n|n-1,i)-1}$ \\
\qquad $	\hat{\bsfg}^{(n|n,i)} = \hat{\bsfg}^{(n|n-1,i)} + \bsfK^{(n,i)}\left(\bsfy^{(n,i)} - \bm\mu_y^{(n|n-1)}\right)$ \\
\qquad $	\hat{\bsfC}_{\bsfg\bsfg}^{(n|n,i)} = \bsfC_{\bsfg\bsfg}^{(n|n-1,i)} - \bsfK^{(n,i)} \bsfC_{\bsfg\bsfy}^{(n|n-1,i)}$ \\
\State \textbf{PF $\#1$ prediction} \\
 \qquad $\bm \xi^{(n+1|n,i)} \sim p\left(\bm \xi^{(n+1|n)}|\bm \Xi^{(n,i)},\bsfY^{(n)}\right)$ \\
\State \textbf{KF prediction for $1 \leq i \leq N_\text{PF}$} \\
\qquad $\hat{\bsfg}^{(n|n-1,i)} = \bsfR^{(n)} \hat{\bsfg}^{(n-1|n-1)}$ \\
\qquad $\bsfC_{\bsfg\bsfg}^{(n|n-1,i)} = \bsfR^{(n)} \bsfC_{\bsfg\bsfg}^{(n-1|n-1,i)} \bsfR^{(n)*} + \bsfX^{(n)}$ \\
\State \textbf{PF $\#2$ to estimate the channel} \\
\qquad $\hat{\bsfh}^{(n)} = \sum_{j=1}^{N_\text{PF}}\tilde{q}^{(n,j)}\bsfB\left(\bm \xi^{(n|n,j)}\right) \hat{\bsfg}^{(n|n,j)}$
\State \textbf{Reshape vectorized channel estimate to find $\hat{\bsfH}^{(n)}[k]$}\\
\qquad $\hat{\bsfH}^{(n)}[k] = \unvect\{\hat{\bsfh}^{(n)}\}$
\end{algorithmic}
\end{algorithm}

\section{Hybrid Precoder and Combiner Design for Channel Tracking}
\label{sec:Prec_Comb_Tracking}

In this section, we propose a subspace-based technique to design hybrid precoders and combiners during channel tracking. Our design aims at designing hybrid precoders and combiners such that the \textit{subspace distance (chordal distance)} between the final precoder (combiner) and the corresponding right (left) channel's singular subspace is minimum, thereby also maximizing the received $\SNR$. 

In the following subsections, we consider precoder design for a single training step $m$, $1 \leq m \leq \Mtck$ during the $n$-th channel slot. The design of the hybrid combiner  is analogous to the design of the hybrid precoder.

\subsection{Chordal distance minimization via subspace projection}
\label{sec:Subspace_Projection}

Let us consider the channel estimates obtained at the end of the $(n-1)$-th TTI, denoted by $\hat{\bsfH}^{(n-1)}[k]$, $0 \leq k \leq K-1$. Their SVDs are given by $\hat{\bsfH}^{(n-1)}[k] = \hat{\bsfU}_H^{(n-1)}[k] \hat{\bm \Sigma}_H^{(n-1)}[k] \hat{\bsfV}_H^{(n-1)*}[k]$, with $\hat{\bsfU}_H^{(n-1)}[k] \in \mathbb{C}^{\Nr \times \mu_L}$, $\hat{\bm \Sigma}_H^{(n-1)}[k] \in \mathbb{C}^{\mu_L \times \mu_L}$, $\hat{\bsfV}_H^{(n-1)}[k] \in \mathbb{C}^{\Nt \times \mu_L}$. Let us use $\hat{\cal H}_\text{c}^{(n-1)}[k] $ to denote the column space of $\hat{\bsfH}^{(n-1)}[k]$, and define the projection matrix onto $\hat{\cal H}_\text{c}^{(n-1)}[k]$ as $\bsfP_{\hat{\cal H}_\text{c}}^{(n-1)}[k] \in \mathbb{C}^{\Nr \times \Nr}$, given by
\begin{equation}
	\bsfP_{\hat {\cal H}_\text{c}}[k] \triangleq \hat{\bsfU}_H^{(n-1)}[k] \hat{\bsfU}_H^{(n-1)*}[k].
	\label{equation:projection_colspace_hatHk}
\end{equation}
Now, let us consider the $\Nt \times \Lt$ hybrid precoder $\bsfF^{(m,n)}[k]$ to be used during the $m$-th channel tracking frame. Such precoder has an SVD $\bsfF^{(m,n)}[k] = \bsfU_F^{(m,n)}[k] \bm \Sigma_F^{(m,n)}[k] \bsfV_F^{(m,n)*}[k]$, with $\bsfU_F^{(m,n)}[k] \in \mathbb{C}^{\Nt \times \Lt}$, $\bm \Sigma_F^{(m,n)}[k] \in \mathbb{C}^{\Lt \times \Lt}$, and $\bsfV_F^{(m,n)}[k] \in \mathbb{C}^{\Lt \times \Lt}$. Let us also use ${\cal C}_F^{(m,n)}$ to denote the column space of $\bsfF^{(m,n)}[k]$, spanned by the column vectors in $\bsfU_F^{(m)}[k]$. The projection matrix onto the column space of $\bsfF^{(m,n)}[k]$ is denoted by the $\Nt \times \Nt$ matrix $\bsfP_{{\cal C}_F}^{(m,n)}[k] = \bsfU_F^{(m,n)}[k] \bsfU_F^{(m,n)*}[k]$.

 Then, the problem of finding $\bsfF^{(m)}[k]$ that minimizes the average chordal distance between ${\cal C}_F^{(m,n)}$ and $\hat{\cal H}_\text{c}^{(n-1)}[k]$ can be stated using the definition of chordal distance as
 \begin{eqnarray}
\lefteqn{\!\!\!\!\!\!\!\!\! 
	\underset{\bsfF^{(m)}[k]}{\min}\,\mathbb{E}_k\left\{\left\|\bsfP_{{\cal C}_F}^{(m,n)}[k] - \bsfP_{\hat {\cal H}_\text{c}}^{(n-1)}[k]\right\|_F^2\right\}}, \label{equation:opt_problem_precoder_tracking} \\
	\mbox{subject to} && \left\{\begin{array}{cc} \frac{1}{\Ns}\sum_{k=0}^{K-1}\|\bsfF^{(m,n)}[k]\|_F^2 &\leq P_\text{tx} \nonumber. \\
	 |[\bsfF_\text{RF}^{(m,n)}]|_{i,j} &= 1, \\ \end{array}\right.
\end{eqnarray}
for $1 \leq i \leq \Nt, 1 \leq j \leq \Lt$. The problem in \eqref{equation:opt_problem_precoder_tracking} is non-convex due to the hardware constraints of the analog precoder \cite{ComaRodGonCasHea:JSTSP:18}, \cite{ElAyach:Spatially_Sparse_Precoding:2014}. Thus, we will neglect their effect momentarily and find the all-digital matrix $\bsfF^{(m,n)}[k]$ that optimizes \eqref{equation:opt_problem_precoder_tracking} to gain some insight into the unconstrained optimum solution. Since we consider frequency-flat precoders and combiners during channel tracking to enable low-complexity channel reconstruction at the receiver side, as explained earlier in this manuscript. we will take $\bsfF^{(m,n)}[k] = \bsfF^{(m,n)}$, for all subcarriers $0 \leq k \leq K-1$. Under this constraint, the cost function in \eqref{equation:opt_problem_precoder_tracking} can be seen to yield

\begin{equation}
	\mathbb{E}_k\left\{\left\|\bsfP_{\hat {\cal H}_\text{c}}^{(n-1)}[k] - \bsfP_{{\cal F}}^{(m)}\right\|_F^2\right\} {\propto} -\underbrace{\sum_{k=0}^{K-1}\left\|\hat{\bsfU}_H^{(n-1)*}[k] \bsfU_F^{(m,n)}\right\|_F^2}_{\gamma^{(m,n)}},
\label{equation:Frob_inner_prod}
\end{equation}
which follows from the definition of chordal distance. Therefore, we are left with the problem of maximizing the only term in \eqref{equation:Frob_inner_prod} that depends on $\bsfU_F$. Let $\hat{\bsfU}_H^{(n-1)} = \sum_{k=0}^{K-1}\hat{\bsfU}_H^{(n-1)}[k] \hat{\bsfU}_H^{(n-1)*}[k]$. Then, the term $\gamma^{(m,n)}$ in \eqref{equation:Frob_inner_prod} yields
\begin{equation}
\begin{split}
	\gamma^{(m,n)} &= \trace\left\{\bsfU_F^{(m,n)*} \hat{\bsfU}_H^{(n-1)}\bsfU_F^{(m,n)}\right\} \\
	&\leq\trace\left\{[\hat{\bm \Lambda}_U^{(n-1)}]_{1:\Lt,1:\Lt}\right\}
\end{split}
\label{equation:upper_bound_Frob}
\end{equation}
where the upper bound in \eqref{equation:upper_bound_Frob} follows from using Von Neumann's trace inequality \cite{VonNeumann}, and it is achieved by setting $\bsfU_F^{(m,n)}$ to the first $\Lt$ eigenvectors of $\hat{\bsfU}_H^{(n-1)}$. Further, since the precoder optimizing \eqref{equation:Frob_inner_prod} only depends on the semi-unitary matrix $\bsfU_F^{(m,n)}$, the all-digital precoding matrix that minimizes the average chordal distance with respect to the estimated channel's row space is $\bsfU_F^{(m,n)}$ itself. Notice, however, that owing to lack of knowledge of $\bsfH^{(n)}[k]$, this choice of $\bsfU_F^{(m,n)}$ does not ensure that the average subspace distance between ${\cal C}_{F}$ and the column space of $\bsfH^{(n)}[k]$ is minimized. If, however, the mmWave channel parameters vary smoothly, then our proposed design criterion approximately holds for $\bsfH^{(n)}[k]$ as well.

Finally, the designed precoder $\bsfU_F^{(m,n)}$ is factorized into an analog $\bsfF_\text{RF}^{(m,n)} \in \mathbb{C}^{\Nt \times \Lt}$ and digital precoder $\bsfF_\text{BB}^{(m,n)} \in \mathbb{C}^{\Lt \times \Ns}$, such that a realizable hybrid precoder $\bsfF_\text{tr}^{(m,n)} = \bsfF_\text{RF}^{(m,n)} \bsfF_\text{BB}^{(m,n)}$ is obtained. The hybrid factorization is performed with the algorithm in \cite{GHP_SPAWC_2015}, which has been shown to provide very good performance and low computational complexity. The design methodology for the hybrid combiner is analogous to that of the hybrid precoder, and we will thus skip it for brevity. Last, we prove that the proposed precoder and combiner design method also maximizes the received $\SNR$ and, consequently, the Fisher Information on the \ac{AoA} and \ac{AoD}.

\textbf{Lemma 1.} Let us consider the SVD of the MIMO channel $\bsfH^{(m,n)}[k] = \bsfU_H^{(m,n)}[k] \bm \Sigma_H^{(m,n)}[k] \bsfV_H^{(m,n)*}[k]$, with $\bsfU_H^{(m,n)}[k] \in \mathbb{C}^{\Nr \times \rank\{\bsfH^{(m,n)}[k]\}}$, $\bsfV_H^{(m,n)}[k] \in \mathbb{C}^{\Nt \times \rank\{\bsfH^{(m,n)}[k]\}}$, $\bm \Sigma_H^{(m,n)}[k] \in \mathbb{C}^{\rank\{\bsfH^{(m,n)}[k]\} \times \rank\{\bsfH^{(m,n)}[k]\}}$. Let us also define two matrices $\bsfG^{(m,n)} \in \mathbb{C}^{\Nr \times \Nr}$ and $\bsfC^{(m,n)} \in \mathbb{C}^{\Nt \times \Nt}$, as
\begin{equation}
\begin{split}
\bsfG^{(m,n)} &= \sum_{k=0}^{K-1} \underbrace{\bsfH^{(m,n)}[k] \bsfF_\text{tr}^{(m,n)} \bsfF_\text{tr}^{(m,n)*} \bsfH^{(m,n)*}[k]}_{\bsfG^{(m,n)}[k]} \\
\bsfC^{(m,n)} &= \underbrace{\sum_{k=0}^{K-1}\bsfH^{(m,n)*}[k] \bsfH^{(m,n)}[k]}_{\bsfC^{(m,n)}[k]}.
\end{split}
\end{equation}
 Then, the all-digital precoder and combiner that maximize the received $\SNR$ are given by
\begin{equation}
	\bsfF_\text{tr}^{(m,n)} = \tilde{\bsfU}_C^{(m,n)} \bsfQ_\text{T} \qquad \bsfW_\text{tr}^{(m,n)} = \tilde{\bsfU}_G^{(m,n)} \bsfQ_\text{R},
\end{equation}
for any arbitrary unitary matrices $\bsfQ_\text{T} \in \mathbb{C}^{\Lt \times \Lt}$, $\bsfQ_\text{R} \in \mathbb{C}^{\Lr \times \Lr}$.

The proof for this Lemma can be found in Appendix B.



\section{Numerical Results}\label{sec:Results}
This section includes the main empirical results obtained with the proposed channel tracking algorithms, \ac{SS-SIGW-OLS} and \ac{GMPF}, and the proposed hybrid precoding and combining algorithm for channel tracking. To obtain these results, we perform Monte Carlo simulations averaged over $100$ trials to evaluate our performance metric, the ergodic rate, as a function of different system parameters. 

The parameters used to configure the transmitter and the receiver in these simulations are summarized as follows. Both the transmitter and the receiver are assumed to be equipped with Uniform Linear Arrays (ULAs) with half-wavelength separation. Such a ULA has steering vectors obeying the expressions $\{\bsfa_\text{T}(\theta_\ell)\}_n = \sqrt{\frac{1}{\Nt}}e^{\sfj n \pi \cos(\theta_\ell)}$, $n = 0,\ldots,\Nt - 1$, and $\{\bsfa_\text{R}(\phi_\ell)\}_m = \sqrt{\frac{1}{\Nr}}e^{\sfj m \pi \cos(\phi_\ell)}$, $m = 0,\ldots,\Nr - 1$. We take $\Nt = 64$, $\Nr = 32$ for illustration. The phase-shifters employed in both transmitter and receiver are assumed to have $N_{\text{Q,Tx}}$ and $N_{\text{Q,Rx}}$ quantization bits, so that the entries of the analog training precoders and combiners are drawn from the sets ${\cal A}_\text{Tx} = \left\{0,\frac{2\pi}{2^{N_{\text{Q,Tx}}}},\ldots,\frac{2\pi (2^{N_{\text{Q,Tx}}}-1)}{2^{N_{\text{Q,Tx}}}}\right\}$ and ${\cal A}_\text{Rx} = \left\{0,\frac{2\pi}{2^{N_{\text{Q,Rx}}}},\ldots,\frac{2\pi (2^{N_{\text{Q,Rx}}}-1)}{2^{N_{\text{Q,Rx}}}}\right\}$. The number of quantization bits is set to $N_\text{Q,Tx} = N_\text{Q,Rx} = 4$ for illustration. The number of RF chains is set to $\Lt = \Lr = 4$. The number of OFDM subcarriers is set to $K = 256$, and a zero-prefix (ZP) length of $Z_\text{P} = K/4 = 64$ samples is assumed to remove Inter Symbol Interference (ISI). The carrier frequency is set to $f_\text{c} = 60$ GHz, the bandwidth is set to $B = 2.55$ GHz, with a roll-off factor of $0.3$ for the raised cosine pulse shaping factor. We use the sampling period defined in the 5G NR standard,  $\Ts = 0.509$ ns.The OFDM symbol duration is $(K + Z_\text{p})\Ts = 0.16$ $\mu$s, and the subcarrier spacing is $1/(K\Ts) = 7.6$ MHz. The transmitted power is set to $P_\text{tx} = 35$ dBm for illustration. 

We generate mmWave frequency-selective channel samples according to \eqref{equation:channel_delay_tap} using \ac{SSP} directly obtained from QuaDRiGa channel simulator \cite{QuaDRiGa_IEEE}, \cite{QuaDRiGa_Tech_Rep} using the scenario 3GPP 38.901 Urban Macrocell (UMa), as described in \cite{5G_channel_model}. We consider two different communication scenarios with different Rician factor and distance between transmitter and receiver, and relative velocity of $v = 300$ km/h $=83.33$ m/s is chosen to evaluate the robustness of the proposed algorithms to high mobility. The scenarios are defined as:
\begin{itemize}
	\item System I: The distance between TX and RX is set to $d = 60$ m. The channel's average Rician factor is set to $K = -7$ dB.
	\item System II: The distance between TX and RX is set to $d = 20$ m. The channel's average Rician factor is set to $K = 0$ dB.
\end{itemize}

\subsection{Performance Metrics}

The main performance metric we focus on in this paper is the ergodic spectral efficiency. Consider the SVD $\hat{\bsfH}^{(n)}[k] = \hat{\bsfU}_H^{(n)}[k] \hat{\bm \Sigma}_H^{(n)}[k] \hat{\bsfV}_H^{(n)*}[k]$, and define $\hat{\bsfH}_\text{eff}^{(t(n),n)}[k] \in \mathbb{C}^{\Ns \times \Ns}$, $\hat{\bsfH}_\text{eff}^{(t(n),n)}[k] \triangleq \hat{\bsfU}_H^{(n)*}[k]\bsfH^{(t(n),n)}[k] \hat{\bsfV}_H^{(n)}[k]$. We consider that, after transmission of $M_\text{tck}$ tracking frames, the receiver takes $T_\text{proc}$ seconds to update the channel estimates and design the precoders and combiners for data transmission (we skip the problem of channel feedback owing to space limitation). Since this paper is focused on devising efficient channel tracking strategies with prior angular information, we focus on all-digital precoders and combiners to gain insight into how well the proposed channel tracking methods perform. With a sampling interval $\Ts = 0.509$ ns \cite{5G_standard}, the parameter $t(n)$ lies within the set ${\cal T}^{(n)} \in (n-1)(M_\text{tck} B_\text{data} + \lceil T_\text{proc}/\Ts \rceil) + (M_\text{tck} + \lceil T_\text{proc}/\Ts \rceil) + 1,\ldots,B_\text{data}$. Let $\eta^{(n)}$ quantify the performance loss owing to sending $M^{(n)}$ frames to track the \ac{mmWave} channel. This quantity is given by $\eta^{(n)} = \Ts B_\text{data}/(\Ts B_\text{data} + T_\text{proc} + \Ts M^{(n)}$. Then, we define the sample average ergodic spectral efficiency as
\begin{equation}
{\cal R}^{(t(n))} = \eta^{(n)} \frac{1}{K}\sum_{k=0}^{K-1}\log_2\left|\bI_{\Ns} + \frac{\SNR}{\Ns} \hat{\bsfH}_\text{eff}^{(t(n))}[k] \hat{\bsfH}_\text{eff}^{(t(n))}[k] \right|.
\label{equation:spectral_efficiency}
\end{equation}


\subsection{Spectral Efficiency Performance}
We evaluate now the spectral efficiency performance of the proposed algorithms for the two \ac{mmWave} scenarios previously defined. For initial channel estimation, we consider our previously proposed \ac{SS-SW-OMP+Th} algorithm, which has been shown to provide near-optimum data rates \cite{RodGonVenHea:TWC_18}.

\begin{figure}[ht!]
\begin{tabular}{cccc}
\hspace*{-4mm}{\includegraphics[width=0.55\columnwidth]{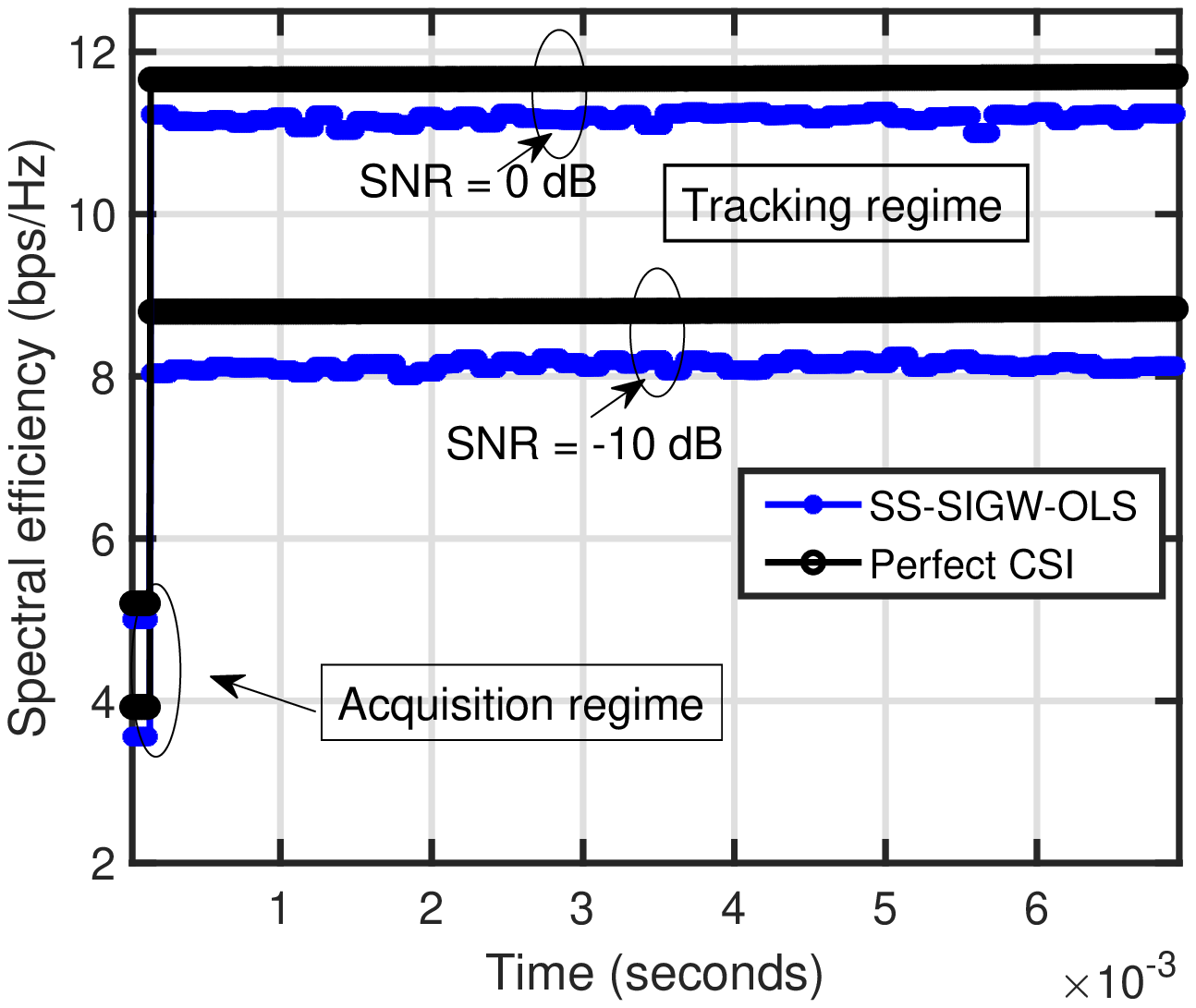}} & 
\hspace*{-8mm}{\includegraphics[width=0.56\columnwidth]{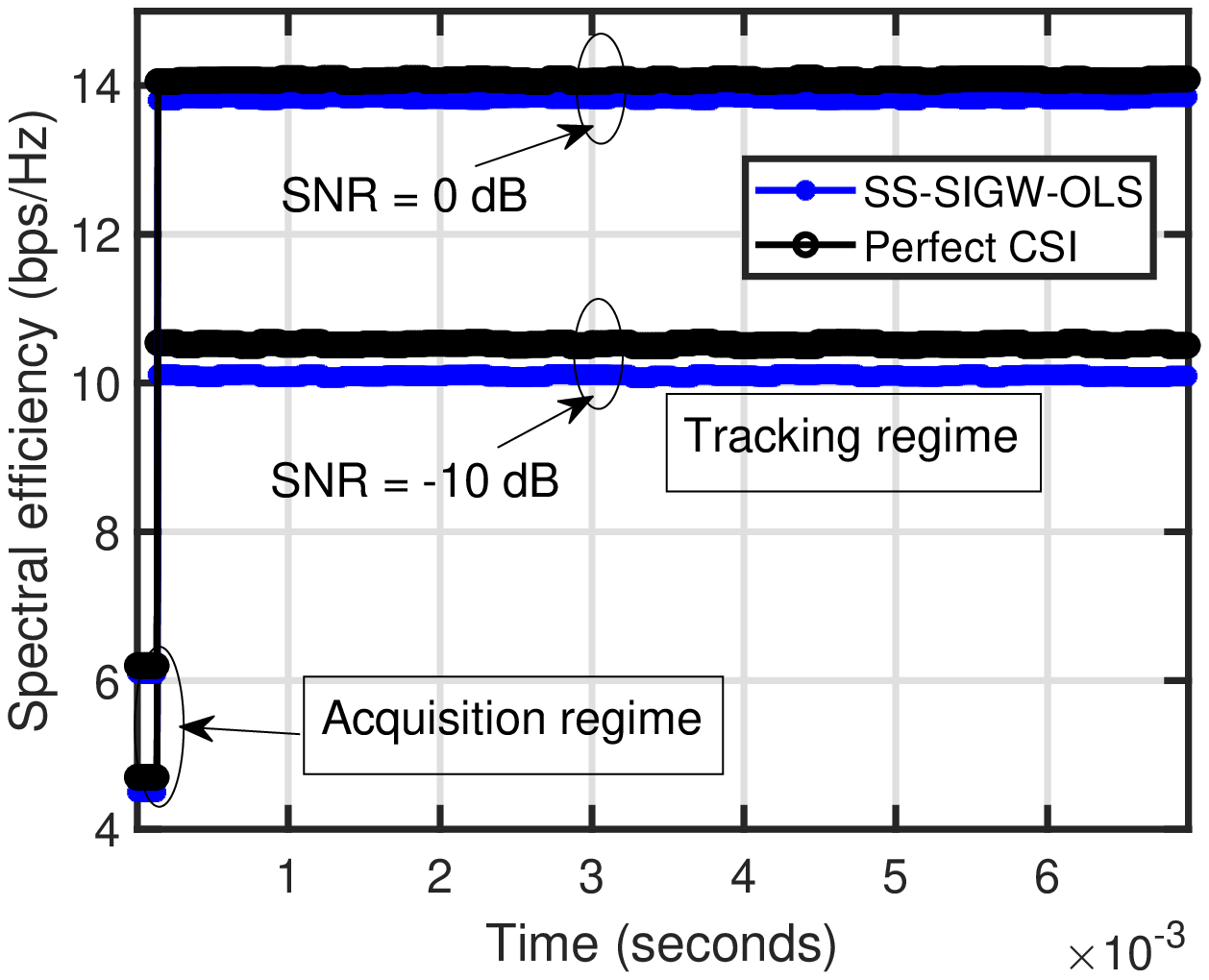}} \\ (a) & (b) \\ 
\hspace*{-4mm} {\includegraphics[width=0.55\columnwidth]{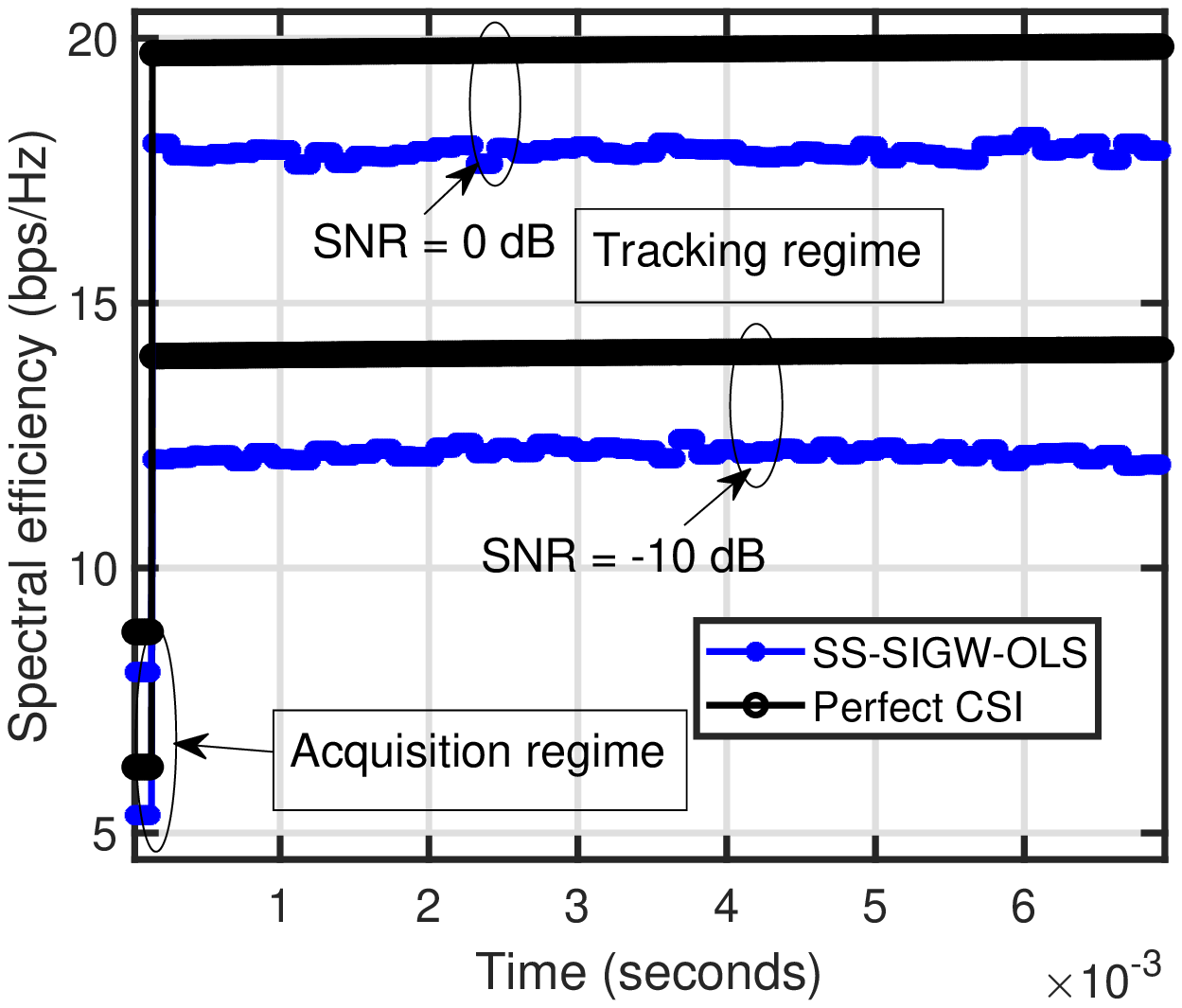}} &
\hspace*{-8mm}{\includegraphics[width=0.56\columnwidth]{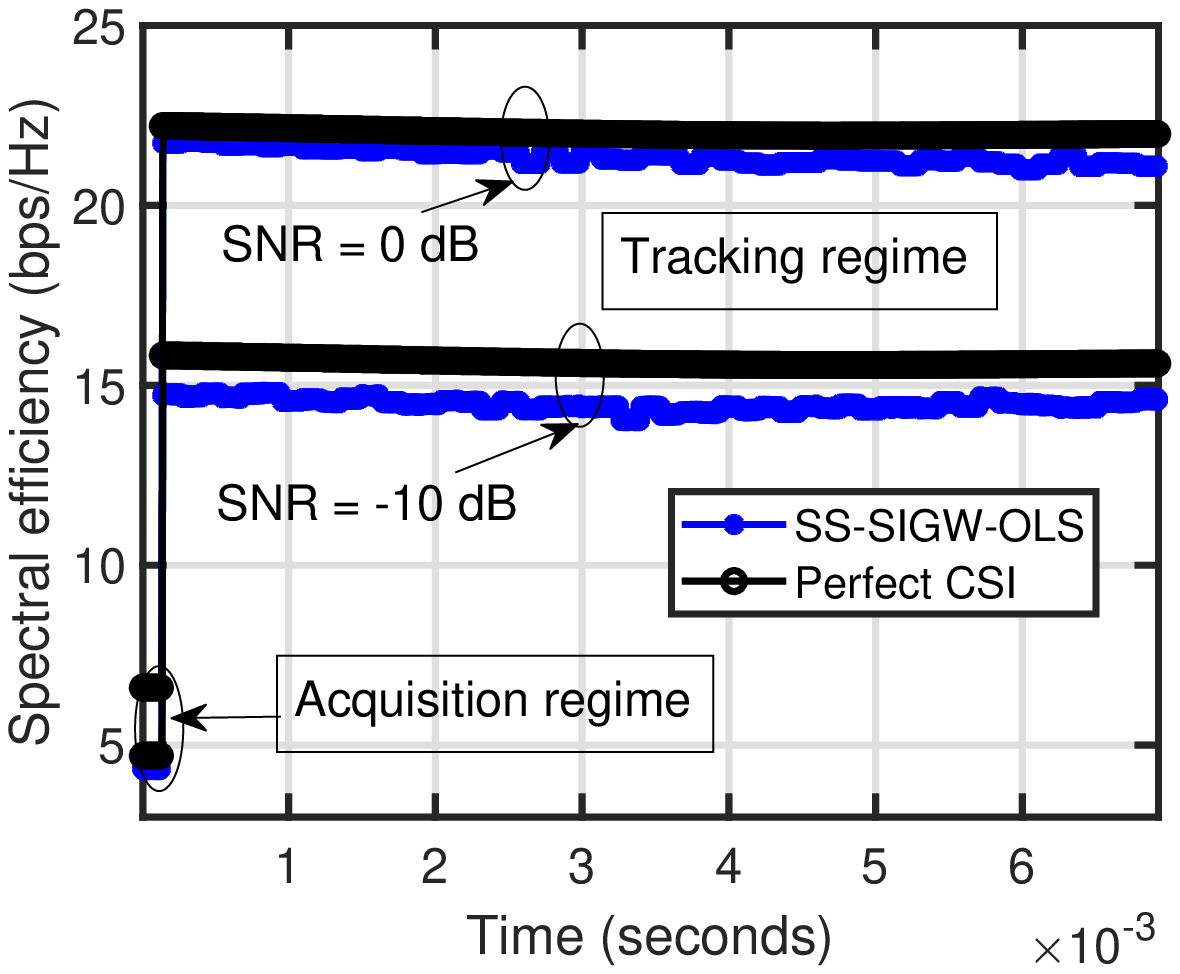}} \\ (c) & (d) \\
    \end{tabular}
    \caption{Comparison of evolution of the spectral efficiency versus time for the \ac{SS-SIGW-OLS} algorithm and perfect CSI, under System I ((a) and (c)) and System II ((b) and (d)), and for different $\SNR$. The number of data streams is set to $\Ns = 1$ ((a) and (b)) and $\Ns = 2$ ((c) and (d)).}.
    \label{fig:SE_vs_Time}
\end{figure}

We show in Fig. \ref{fig:SE_vs_Time} the evolution of the spectral efficiency as a function of time for both System I and System II, for $\SNR = \{-10,0\}$ dB, and $\Ns = \{1,2\}$. The number of training frames for initial estimation is set to $M_\text{ini} = 80$, and the number of tracking frames is set to $M_\text{tck} = 8$, for illustration. Each block of training symbols plus data is assumed to have fixed length, and consists of $240$ OFDM symbols. The number of subcarriers processed by our proposed \ac{SS-SIGW-OLS} algorithm is set to $K_\text{p} = 64$, and the threshold parameter is set to $0.025 \sigma^2$. In Fig. \ref{fig:SE_vs_Time}, we observe two different regimes, namely \textit{acquisition regime} and \textit{tracking regime}. For the former, the spectral efficiency achieved is not high, which comes from the different penalty factor $\eta^{(n)}$. For the latter, the spectral efficiency exhibits a small gap with respect to perfect CSI, and this gap increases with $\Ns$, owing to the difficulty in estimating the left and right singular basis of the \ac{mmWave} MIMO channel. During initial acquisition, the number of data symbols is $B_\text{data} = 240 - M_\text{ini} = 186$ OFDM symbols, while during tracking this number is $B_\text{data} = 240 - M_\text{tck} = 232$ OFDM symbols. Further, the receive $\SNR$ during initial acquisition is very low, while it increases by a factor proportional to $\Nt \Nr$ during tracking. {\em Near optimal data rates are obtained even is the system is being tested with channel series generated from Quadriga, that vary at the symbol rate, that is, the channel is not really static during the channel slot, as assumed for the mathematical derivation of the algorithms}. For System I, the gap between the achieved performance and the one obtained with perfect CSI is larger, due to the smaller Rician factor, that leads to  more multiptah components to be estimated and tracked.


\subsection{Tracking overhead performance}
\begin{figure}[ht!]
\begin{tabular}{cccc}
\hspace*{-4mm}{\includegraphics[width=0.55\columnwidth]{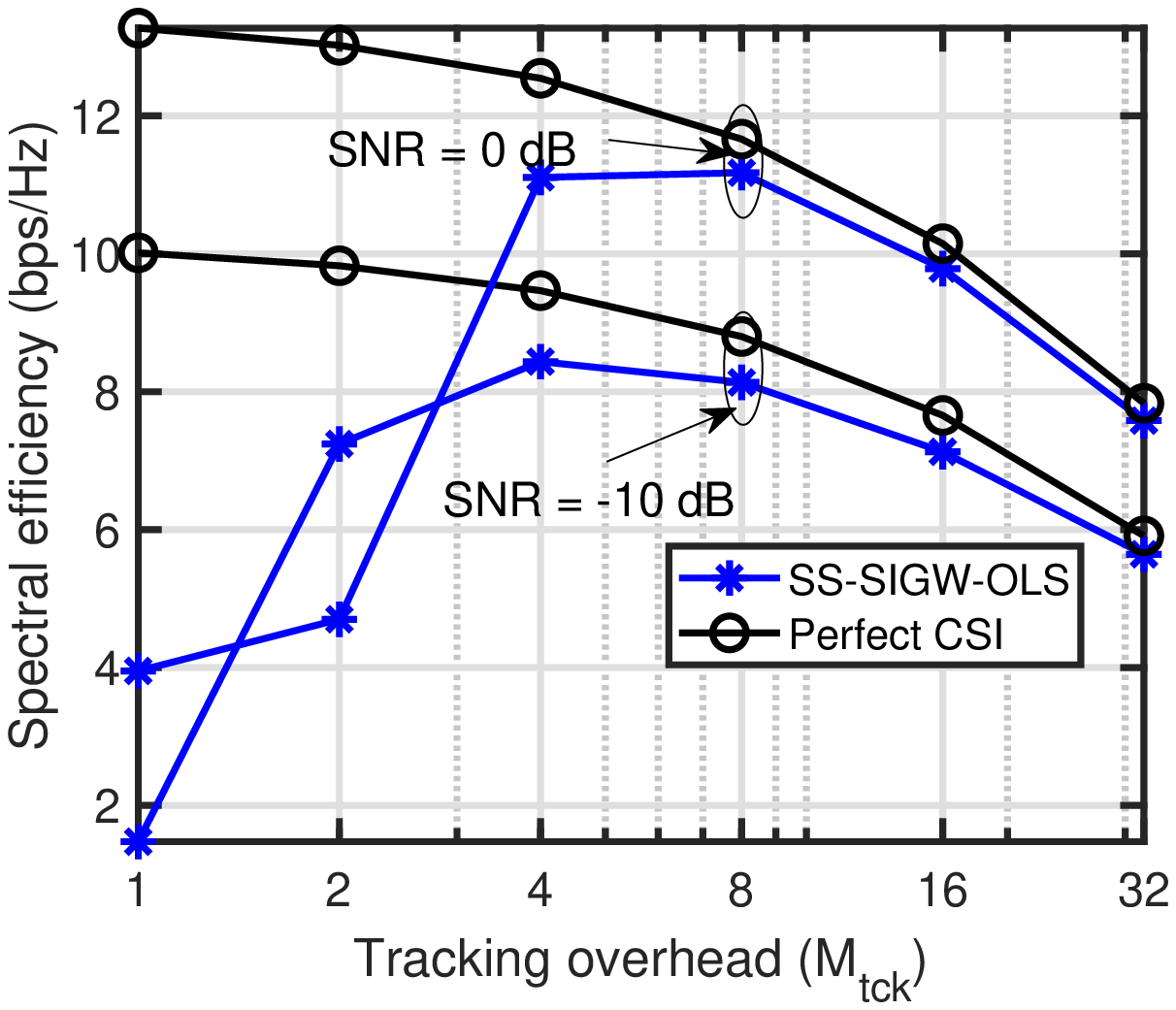}} & \hspace*{-8mm}{\includegraphics[width=0.55\columnwidth]{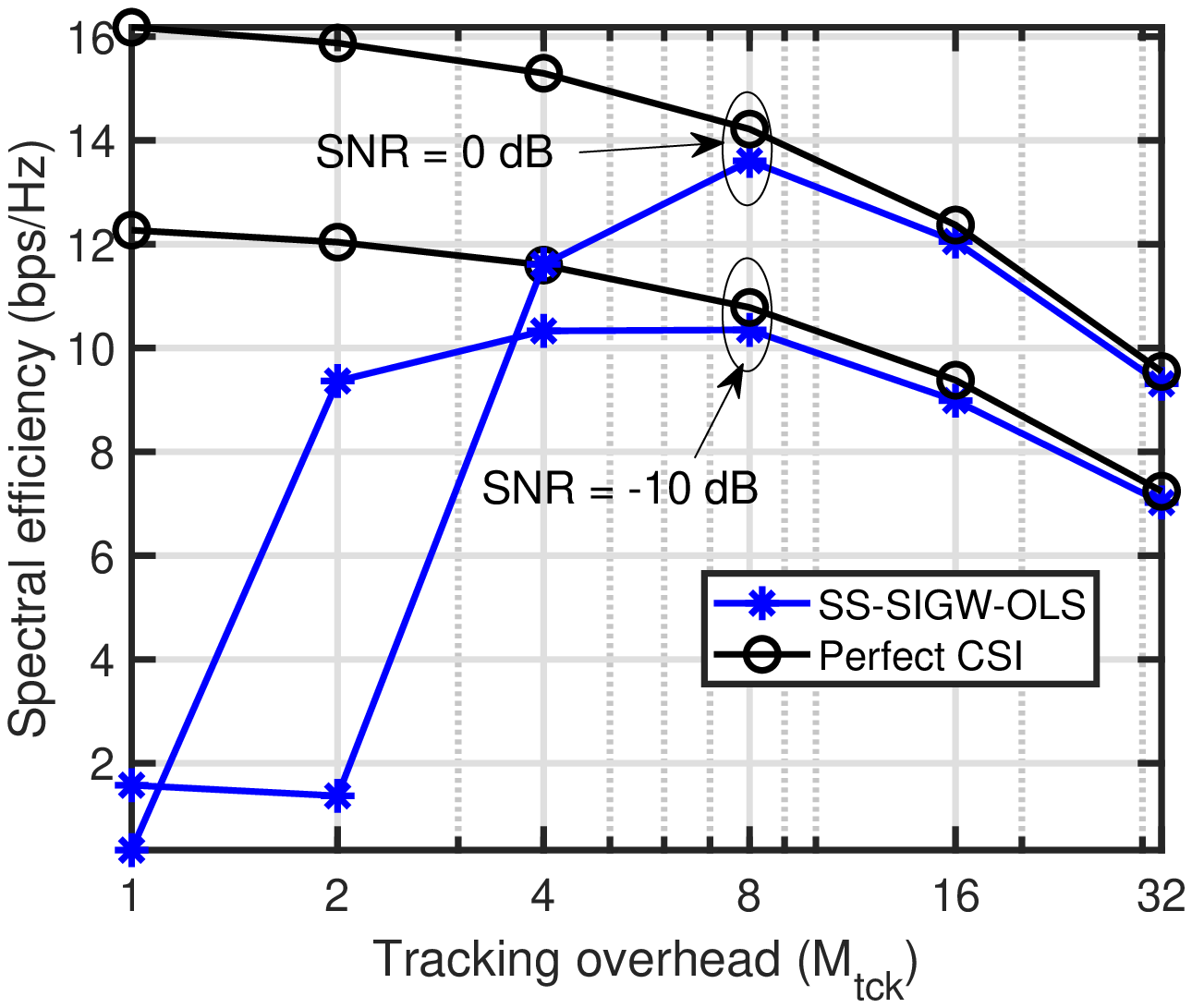}} \\ (a) & (b) \\
\hspace*{-4mm}{\includegraphics[width=0.56\columnwidth]{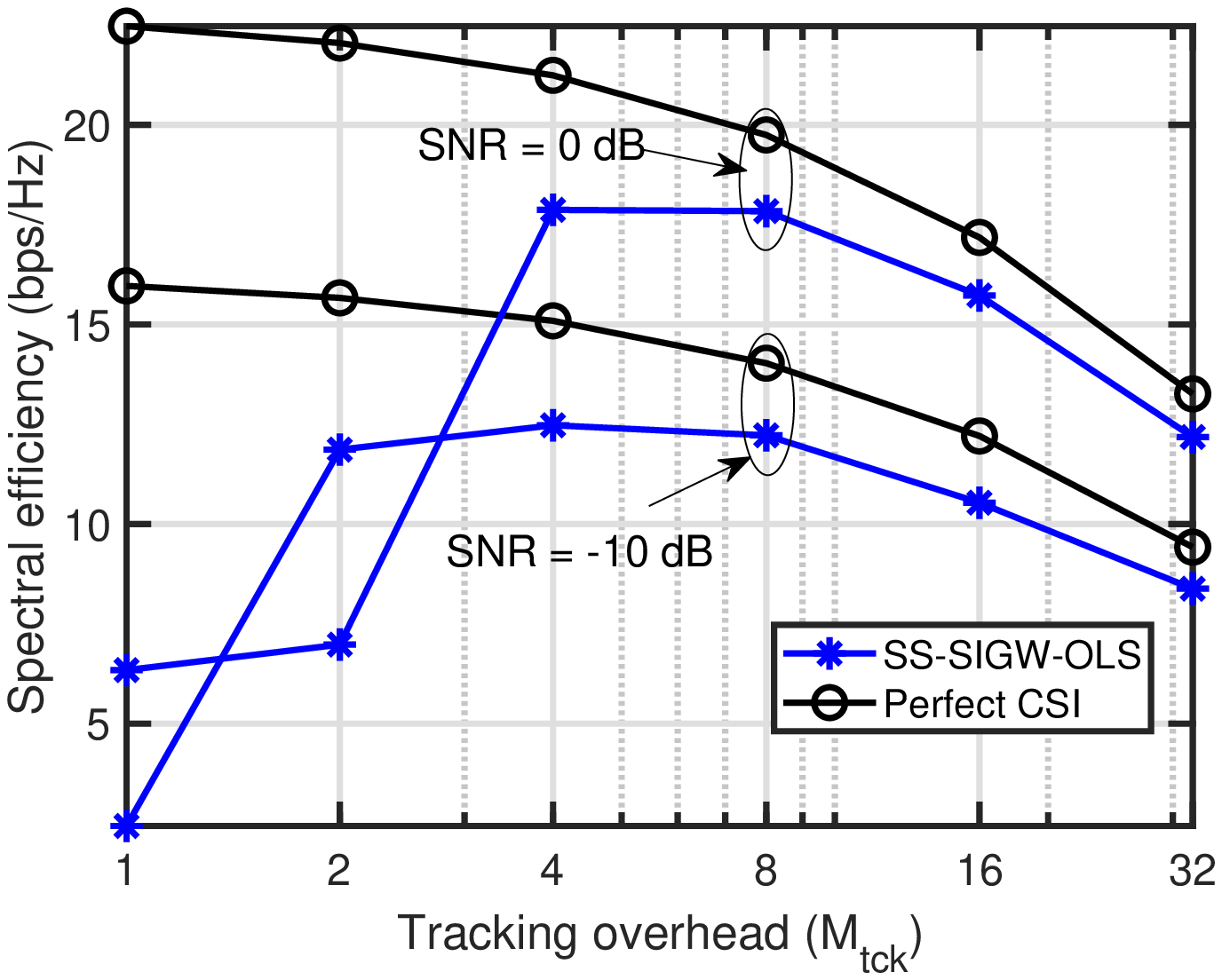}} &  \hspace*{-8mm} {\includegraphics[width=0.54\columnwidth]{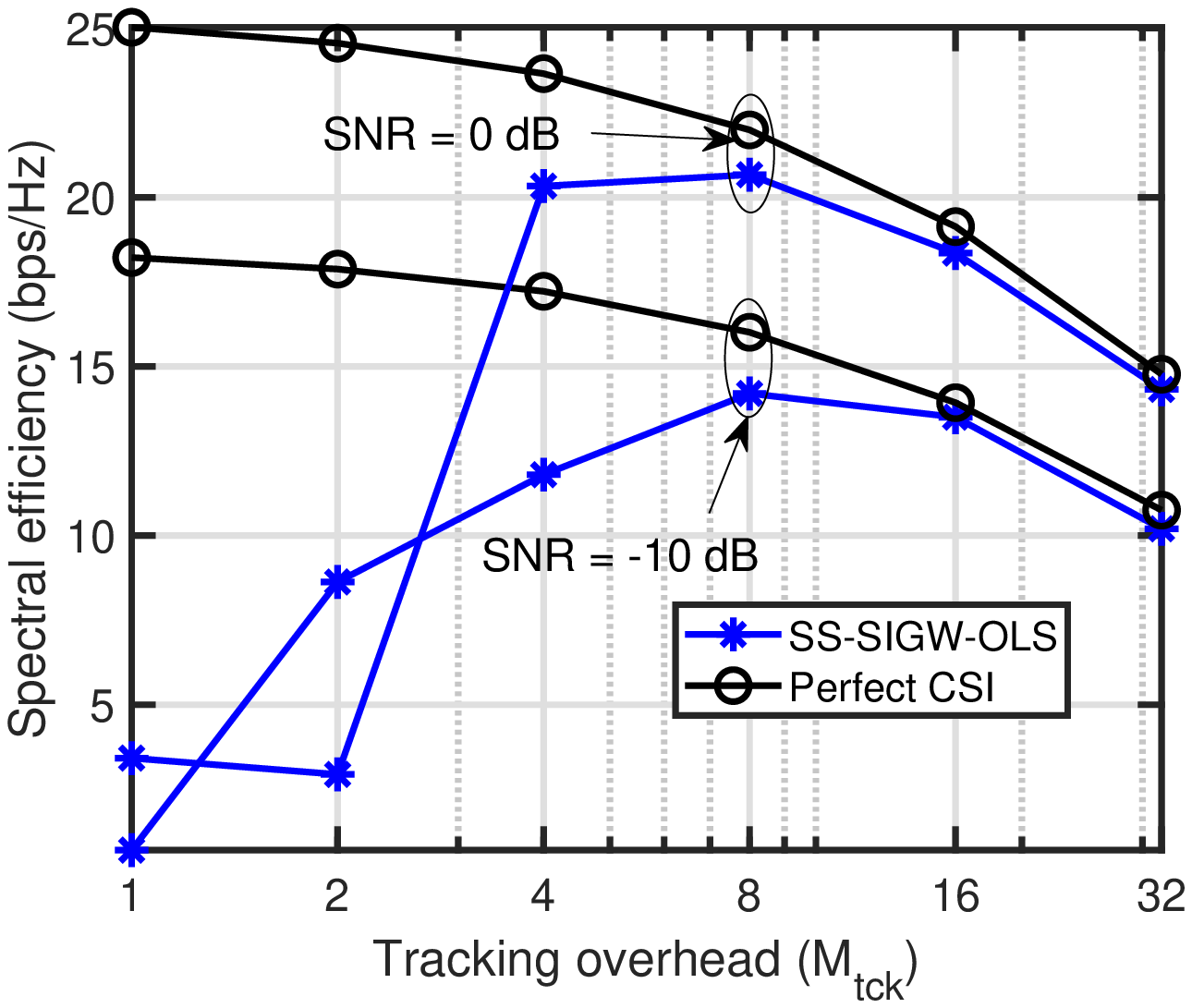}} \\
(c) & (d) \\
    \end{tabular}
    \caption{Comparison of evolution of the spectral efficiency versus time for the \ac{SS-SIGW-OLS} algorithm and perfect CSI, under System I ((a) and (c)) and System II ((b) and (d)), and for different $\SNR$. The number of data streams is set to $\Ns = 1$ ((a) and (b)) and $\Ns = 2$ ((c) and (d)).}.
    \label{fig:SE_vs_Overhead}
\end{figure}

In this subsection, we study the spectral efficiency performance as a function of the overhead of the proposed algorithms. We analyze first the performance of  \ac{SS-SIGW-OLS} for both System I and System II, for $\SNR = \{-10,0\}$ dB, and $\Ns = \{1,2\}$. We show in Fig. \ref{fig:SE_vs_Overhead}, the ergodic spectral efficiency, averaged over $100$ channel slots, versus the tracking overhead $M_\text{tck}$, which ranges from $1$ up to $32$ OFDM symbols. We observe that the spectral efficiency of \ac{SS-SIGW-OLS} increases with $M_\text{tck}$ until it reaches a maximum, and then decreases. This comes from the penalty factor $\eta^{(n)}$ we introduced in \eqref{equation:spectral_efficiency}. As $M_\text{tck}$ increases, the penalty factor decreases, but our proposed algorithm is able to find better channel estimates, such that higher spectral efficiency values are attained, reaching a maximum around $M_\text{tck} = 8$. {\em When $M_\text{tck}$ is larger than a threshold, however, the performance improvement coming from having better channel estimates does not compensate for the additional number of training pilots that have to be sent}.  We also observe that the performance gap for System II is smaller than that of System I, which comes again from the larger Rician factor in System II.

\begin{figure}[ht!]
\begin{tabular}{cccc}
\hspace*{-4mm}{\includegraphics[width=0.55\columnwidth]{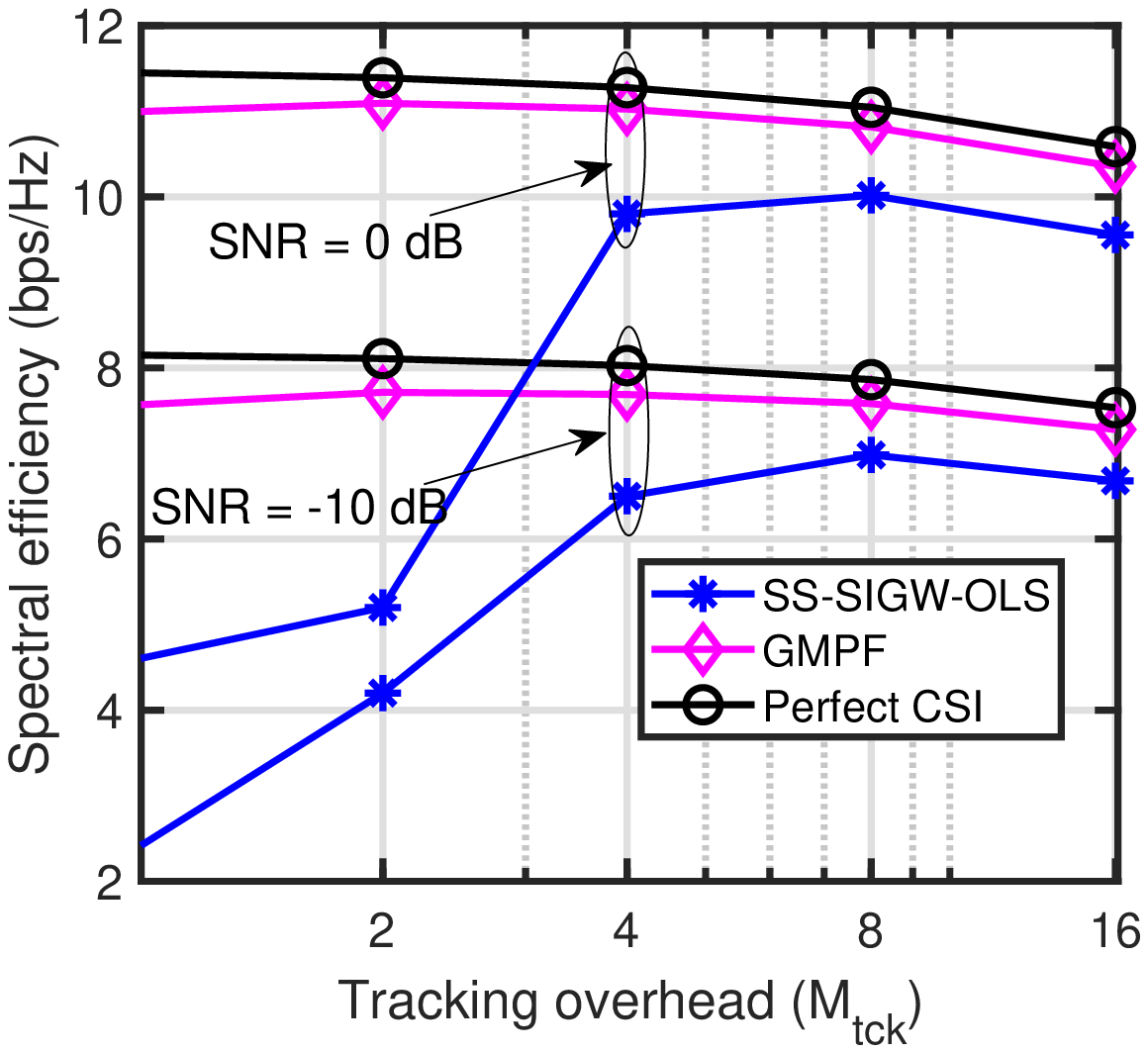}} & \hspace*{-7mm} {\includegraphics[width=0.55\columnwidth]{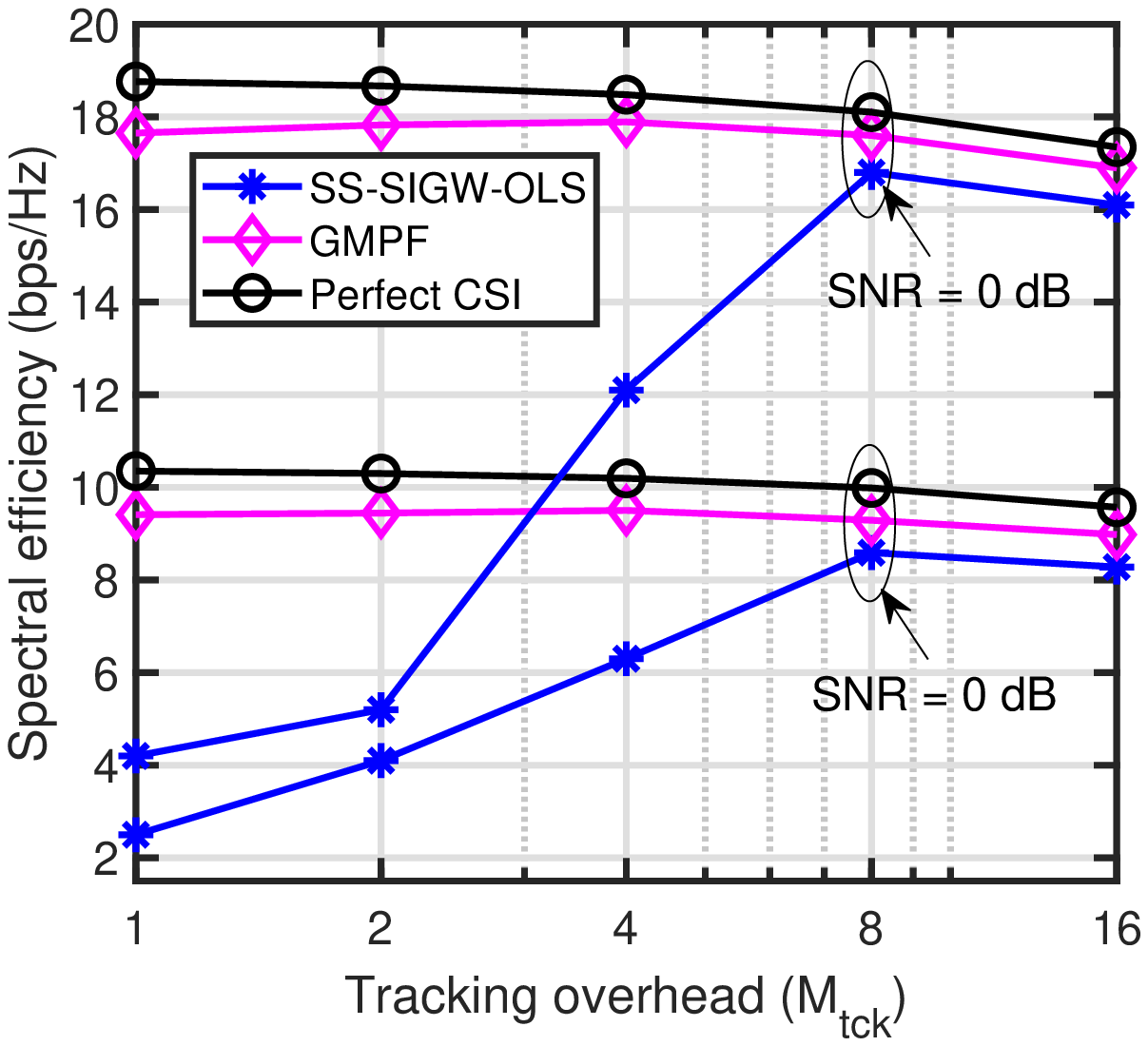}} \\
(a) & (b) \\
    \end{tabular}
    \caption{Comparison of evolution of the spectral efficiency versus time for the \ac{SS-SIGW-OLS} and \ac{GMPF} algorithms and perfect CSI for different $\SNR$. The number of data streams is set to $\Ns = 1$ (a) and $\Ns = 2$ (b).}.
    \label{fig:Bayesian_vs_classical_overhead}
\end{figure}

Finally, we show in Fig. \ref{fig:Bayesian_vs_classical_overhead} the spectral efficiency performance as a function of overhead for both the proposed \ac{SS-SIGW-OLS} and \ac{GMPF}, for $\SNR = \{-10,0\}$ dB, and $\Ns = \{1,2\}$. In these simulations, we consider the channel has a \ac{LOS} component with a single ray \cite{5G_channel_model}, and $C = 2$ \ac{NLOS} clusters, each contributing with a number of rays $R_c \sim {\cal U}[6,20]$. The system parameters are the same as those in Fig. \ref{fig:SE_vs_Overhead}. We consider a relative velocity of $\pm 30$ m/s for each component of the velocity vector of the user, and the average Rician factor is set to $K_\text{factor} = 0$ dB. The proposed \ac{GMPF} algorithm is configured to use $N_\text{PF} = 1000$ particles to discretize the joint Laplacian pdf of the \ac{AoA}/\ac{AoD}. The correlation and covariance matrices $\bsfR^{(n)}$, $\bsfX^{(n)}$ in \eqref{equation:evolution_channel_gains} are taken from \cite{ICC_2019_ERBPF}, \cite{Rap:17:Spatial_correlation}.  In Fig. \ref{fig:Bayesian_vs_classical_overhead}, we observe the additional benefit coming from exploiting prior statistical information, whereby the \ac{GMPF} algorithm clearly outperforms its classical counterpart. It is important to highlight that this performance improvement comes at the expense of higher computational complexity, since $N_\text{PF}$ particles are used to estimate the \ac{AoA}, \ac{AoD}, channel gains, and the vectorized frequency-selective channel. Further, the gain coming from exploiting prior statistical information on the \ac{AoA}, \ac{AoD}, and channel gains, is also reflected in the required training overhead to achieve a given target spectral efficiency value. {\em Our classical \ac{SS-SIGW-OLS} algorithm requires around $M_\text{tck} = 8$ tracking frames, while our proposed Bayesian \ac{GMPF} algorithm only requires $M_\text{tck} = 1$ tracking frame to estimate the channel}, which helps further reduce computational complexity.

\section{Conclusions}\label{sec:Conclusion}

In this paper, we formulated the problem of channel tracking for a MIMO-OFDM system operating at \ac{mmWave}, and proposed both a \textit{classical} sparsity-constrained \ac{ML} estimator, and a non-linear \textit{Bayesian} filter aiming at finding the global \ac{MMSE} estimator. We also proposed an SNR-maximizing precoding and combining design method for channel tracking, leveraging prior information on the previously estimated \ac{AoA} and \ac{AoD}, such that low-overhead channel tracking can be performed efficiently. Simulation results showed the effectiveness of our proposed methods, which have been evaluated using realistic channel realizations extracted from the 5G \ac{NR} channel model. We showed that, even under high mobility conditions, the proposed tracking framework is able to maintain near-optimum spectral efficiency values at low overhead, even if the distance between transmitter and receiver is small. Future work would conduct the extension of the proposed framework to deal with channel blockage, and consider a multi-user scenario, without assuming perfect synchronization, and accounting for the beam-squint effect that is present when the signal bandwidth is very large.

\section*{Appendix A: Derivation of the gradient vectors}\label{sec:appendixA}

In this appendix, the explicit derivation of the partial derivatives of $\bP(\bm \theta,\bm \phi)$ with respect to $\theta_\ell$ and $\phi_\ell$, $l = 1,\ldots,L$ is presented. We will focus on the $\ell$-th AoD $\theta_\ell$ for illustration, since the explicit calculation is the same for every parameter. Then, the derivative of the projection matrix in \eqref{equation:optimization_gridless} can be found as
\begin{equation}
\begin{split}
	\frac{\partial \bP(\bm \theta,\bm \phi)}{\partial \theta_\ell} =& \bD_\text{w}^{-*}\frac{\partial}{\partial \theta_\ell}{\bigg{[}}\bm \Upsilon(\bm \theta,\bm \phi)(\bm \Upsilon^*(\bm \theta,\bm \phi)\bC_\text{w}^{-1}\bm \Upsilon(\bm \theta,\bm \phi))^{-1} \\ &\times \bm \Upsilon^*(\bm \theta,\bm \phi){\bigg{]}}\bD_\text{w}^{-1}.
\end{split}
\end{equation}
Now, let us define a matrix $\bm \Gamma(\bm \theta,\bm \phi) \in \mathbb{C}^{L \times ML_\text{r}}$, $\bm \Gamma(\bm \theta,\bm \phi) = (\bm \Upsilon^*(\bm \theta,\bm \phi)\bC_\text{w}^{-1}\bm \Upsilon(\bm \theta,\bm \phi))^{-1}\bm \Upsilon^*(\bm \theta,\bm \phi)$. Thereby, the derivative of the term within brackets is given by
\begin{equation}
	\frac{\partial \bm \Upsilon(\bm \theta,\bm \phi)\bm \Gamma(\bm \theta,\bm \phi)}{\partial \theta_\ell} = \frac{\partial \bm \Upsilon(\bm \theta,\bm \phi)}{\partial \theta_\ell}\bm \Gamma(\bm \theta,\bm \phi) + \bm \Upsilon(\bm \theta,\bm \phi)\frac{\partial \bm \Gamma(\bm \theta,\bm \phi)}{\partial \theta_\ell}.
	\label{equation:primal_derivative}
\end{equation}
The derivative of $\bm \Upsilon(\bm \theta,\bm \phi)$ is given by
\begin{equation}
\begin{split}
	\frac{\partial \bm \Upsilon(\bm \theta,\bm \phi)}{\partial \theta_\ell} &= \bm \Phi \overbrace{\left[\begin{array}{ccccccc}	 
	\bm 0 & \ldots & \bm 0 & \frac{\partial \bm \psi(\theta_\ell,\phi_l)}{\partial \theta_\ell} & \bm 0 & \ldots & \bm 0 \\ 
	\end{array}\right]}^{\frac{\partial \bm \Psi(\bm \theta,\bm \phi)}{\partial \theta_\ell}} \\
	&= \left[\begin{array}{ccccccc} 
	\bm 0 & \ldots & \bm 0 & \bm \Phi \frac{\partial \bm \psi(\theta_\ell,\phi_l)}{\partial \theta_\ell} & \bm 0 & \ldots & \bm 0 \\ \end{array}\right],
\end{split}
\end{equation}	
where the derivative of $\bm \psi(\theta_\ell,\phi_l)$ with respect to $\theta_\ell$ is given by
\begin{equation}
	\frac{\partial \bm \psi(\theta_\ell,\phi_l)}{\partial \theta_\ell} = \frac{\partial {\ba_\text{T}^\text{C}(\theta_\ell)}}{\partial \theta_\ell} \otimes \ba_\text{R}(\phi_l)).
\end{equation}
On the other hand, the derivative of $\bm \Gamma(\bm \theta,\bm \phi)$ can be found to be
\begin{equation}
	\frac{\partial \bm \Gamma(\bm \theta,\bm \phi)}{\partial \theta_\ell} = \frac{\partial \bM^{-1}(\bm \theta,\bm \phi)}{\partial \theta_\ell}\bm \Upsilon^*(\bm \theta,\bm \phi) + \bM^{-1}(\bm \theta,\bm \phi) \frac{\partial \bm \Upsilon^*(\bm \theta,\bm \phi)}{\partial \theta_\ell},
	\label{equation:Gamma_derivative}
\end{equation}
where $\bM(\bm \theta,\bm \phi) \in \mathbb{C}^{L \times L}$ is defined as $\bM(\bm \theta,\bm \phi) = \bm \Upsilon^*(\bm \theta,\bm \phi)\bC_\text{w}^{-1}\bm \Upsilon(\bm \theta,\bm \phi)$. The derivative of $\bM^{-1}(\bm \theta,\bm \phi)$ can be found as
\begin{equation}
	\frac{\partial \bM^{-1}(\bm \theta,\bm \phi)}{\partial \theta_\ell} = -\bM^{-1}(\bm \theta,\bm \phi)\frac{\partial \bM(\bm \theta,\bm \phi)}{\partial \theta_\ell}\bM^{-1}(\bm \theta,\bm \phi),
	\label{equation:derivative_invM}
\end{equation}
where the derivative of $\bM(\bm \theta,\bm \phi)$ is given by
\begin{equation}
	\frac{\partial \bM(\bm \theta,\bm \phi)}{\partial \theta_\ell} = \frac{\partial }{\partial \theta_\ell}{\bigg{[}} \bm \Upsilon^*(\bm \theta,\bm \phi)\bC_\text{w}^{-1}\bm \Upsilon(\bm \theta,\bm \phi){\bigg{]}}.
	\label{equation:derivative_M_1}
\end{equation}
The derivative in \eqref{equation:derivative_M_1} can be found as
\begin{equation}
\begin{split}
	\frac{\partial \bM(\bm \theta,\bm \phi)}{\partial \theta_\ell} =& \bm \Upsilon^*(\bm \theta,\bm \phi)\bC_\text{w}^{-1} \bm \Phi \frac{\partial \bm \Psi(\bm \theta,\bm \phi)}{\partial \theta_\ell} \\&+ \frac{\partial \bm \Psi^*(\bm \theta,\bm \phi)}{\partial \theta_\ell}\bm \Phi^* \bC_\text{w}^{-1}\bm \Upsilon(\bm \theta, \bm \phi).
\end{split}
	\label{equation:derivative_M}
\end{equation}
Finally, plugging \eqref{equation:derivative_M} into \eqref{equation:derivative_invM}, and then in \eqref{equation:Gamma_derivative} yields the final gradient for $\theta_\ell$. The same procedure is followed for calculation of the gradient with the remaining angular parameters. Stacking the different terms in a column vector yieds the final form of the gradient in \eqref{equation:gradient_ML}.

\section*{Appendix B: Proof of Lemma 1}\label{sec:appendixB}

\begin{proof}
Let us consider the eigendecompositions $\bsfG^{(m,n)} = \bsfU_G^{(m,n)} \bm \Lambda_G^{(m,n)} \bsfU_G^{(m,n)*}$, and $\bsfC^{(m,n)} = \bsfU_C^{(m,n)} \bm \Lambda_C^{(m,n)} \bsfU_C^{(m,n)*}$. Then,
\begin{equation}
\begin{split}
	&\sum_{k=0}^{K-1}\|\bsfW_\text{tr}^{(m,n)*} \bsfH^{(m,n)}[k] \bsfF_\text{tr}^{(m,n)}\|_F^2 = \\ & = \trace \left\{\bsfW_\text{tr}^{(m,n)*} \left(\sum_{k=0}^{K-1} \bsfG^{(m,n)}[k] \right) \bsfW_\text{tr}^{(m,n)} \right\} \\
	&= \trace\left\{ \bsfW_\text{tr}^{(m,n)*} \bsfU_G^{(m,n)} \bm \Lambda_G^{(m,n)} \bsfU_G^{(m,n)*} \bsfW_\text{tr}^{(m,n)} \right\}
\\ &\overset{(a)}{\leq} \trace\left\{\tilde{\bsfU}_G^{(m,n)} \tilde{\bm \Lambda}_G^{(m,n)} \tilde{\bsfU}_G^{(m,n)*} \right\} \\
	&= \trace\left\{\bsfF_\text{tr}^{(m,n)*} \left(\sum_{k=0}^{K-1} \bsfH^{(m,n)*}[k] \bsfH^{(m,n)}[k] \right) \bsfF_\text{tr}^{(m,n)}\right\} \\
	&\overset{(b)}{\leq} \trace\left\{\left[\bm \Lambda_C^{(m,n)}\right]_{1:\Lt,1:\Lt}\right\},
\end{split}
\label{proof_SNR_maximization}
\end{equation}
where $(a)$ follows from applying Von Neumann's trace inequality \cite{VonNeumann}, and equality is attained by setting $\bsfW_\text{tr}^{(m,n)} = \left[\bsfU_G^{(m,n)}\right]_{:,1:\Lr} \bsfQ_\text{R}$, for any $\bsfQ_\text{R} \in \mathbb{C}^{\Lr \times \Lr}$ satisfying $\bsfQ_\text{R} \bsfQ_\text{R}^* = \bI_{\Lr}$. Step $(b)$ also follows from using Von Neumann's trace inequality, and equality is achieved if $\bsfF_\text{tr}^{(m,n)} = \left[\bsfU_C^{(m,n)}\right]_{:,1:\Lt} \bsfQ_\text{T}$, for any $\bsfQ_\text{T} \in \mathbb{C}^{\Lt \times \Lt}$ fulfilling $\bsfQ_\text{T} \bsfQ_\text{T}^* = \bI_{\Lt}$. Since the receive all-digital combiner is semi-unitary, it does not alter noise statistics, and maximizing \eqref{proof_SNR_maximization} is equivalent to maximizing the receive $\SNR$. Finally, projecting an input precoder onto the subspace spanned by the eigenvectors of $\bsfG^{(m,n)}$ and $\bsfC^{(m,n)}$ leads to the projected precoder belonging to the same subspace, so that the all-digital corresponding projection matrices are optimal. Consequently, Fisher Information can be also seen to be maximized using the Slepian-Bangs formula. This concludes the proof.
\end{proof}

\bibliographystyle{IEEEtran}


\end{document}